\documentclass[preprint,tightenlines,superscriptaddress,eqsecnum,floats,nofootinbib,showpacs]
{revtex4}

\usepackage{amsmath,amssymb,amsfonts}
\usepackage{graphicx}

\def\be{\begin{equation}}
\def\ee{\end{equation}}
\def\ba{\begin{eqnarray}}
\def\ea{\end{eqnarray}}
\def\nn{\nonumber}

\newcommand{\secref}[1]{Sec.~\ref{#1}}
\newcommand{\eqnref}[1]{(\ref{#1})}
\newcommand{\figref}[1]{Fig.~\ref{#1}}

\newcommand{\Pl}{\ell_\mathrm{Pl}} 
\newcommand{\mubar}{{\bar \mu}} 
\newcommand{\abs}[1]{{\left|{#1}\right|}} 
\newcommand{\pinner}[2]{({#1},{#2})_\mathrm{phy}} 
\newcommand{\inner}[2]{{\langle {#1}\vert {#2} \rangle}} 
\newcommand{\ket}[1]{\vert{#1}\rangle} 

\newcommand{\sgn}{\mathrm{sgn}} 
\newcommand{\Tr}{\mathrm{Tr}} 
\newcommand{\grav}{\mathrm{grav}} 
\newcommand{\matt}{\mathrm{matt}} 
\newcommand{\kin}{\mathrm{kin}} 
\newcommand{\phy}{\mathrm{phy}} 
\newcommand{\hil}{\mathcal{H}} 

\newcommand{\ftriad}[2]{{}^o\! e^{#1}_{#2}} 
\newcommand{\fcotriad}[2]{{}^o\!\omega_{#1}^{#2}} 
\newcommand{\fq}{{}^o\!q} 
\newcommand{\tE}{\mbox{$\tilde{E}$}} 

\begin{document}


\title{Loop quantum cosmology with higher order holonomy corrections}
\author{Dah-Wei Chiou}
\email{chiou@gravity.psu.edu}
\affiliation{
Department of Physics, Beijing Normal University, Beijing 100875, China}
\author{Li-Fang Li}
\email{lilifang@mail.bnu.edu.cn}
\affiliation{
Department of Physics, Beijing Normal University, Beijing 100875, China}

\begin{abstract}
With a well-motivated extension of higher order holonomy corrections, the quantum theory of loop quantum cosmology (LQC) for the $k=0$ Friedmann-Robertson-Walker model (with a free massless scalar) is rigorously formulated. The analytical investigation reveals that, regardless of the order of holonomy corrections and for any arbitrary states, the matter density remains finite, bounded from above by an upper bound, which equals the critical density obtained at the level of heuristic effective dynamics. Particularly, with all orders of corrections included, the dynamical evolution is shown to follow the bouncing scenario in which two Wheeler-DeWitt (WDW) solutions (expanding and contracting) are bridged together through the quantum bounce. These observations provide further evidence that the quantum bounce is essentially a consequence of the intrinsic discreteness of LQC and LQC is fundamentally different from the WDW theory. Meanwhile, the possibility is also explored that the higher order holonomy corrections can be interpreted as a result of admitting generic $SU(2)$ representations for the Hamiltonian constraint operators.
\end{abstract}


\maketitle


\section{Introduction}\label{sec:introduction}
In previous years, the status of loop quantum cosmology (LQC) has undergone lively progress and become an active area of research. Specifically, with the inclusion of a free massless scalar field, the comprehensive formulation of LQC in the $k=0$ Friedmann-Robertson-Walker (FRW) (i.e. spatially flat and isotropic) model has been constructed in detail, providing a solid foundation for the quantum theory and showing that the big bang singularity is resolved and replaced by the \emph{quantum bounce}, which bridges the present universe with a preexisting one \cite{Ashtekar:2006rx,Ashtekar:2006uz,Ashtekar:2006wn}. Similar results are also affirmed for a variety of extended models \cite{Ashtekar:2006es,Vandersloot:2006ws,Chiou:2006qq,Szulc:2008ar}.

However, it remains questionable whether the quantum bounce indeed results intimately from the quantum nature of Riemannian geometry of loop quantum gravity (LQG), as the same result can be readily obtained even at the level of heuristic effective dynamics without appealing to the sophisticated features of LQC \cite{Chiou:2007dn,Chiou:2007sp,Chiou:2007mg}.

In response to this criticism, a simplified but exactly soluble model of LQC has been studied and used to prove that resolution of the classical singularity and occurrence of the quantum bounce are robust \cite{Ashtekar:2007em}. Furthermore, the study of \cite{Ashtekar:2007em} brings out the precise sense in which the Wheeler-DeWitt (WDW) theory approximates LQC and the sense in which this approximation fails, thereby showing that LQC is intrinsically discrete and the underlying discreteness is essential for the quantum bounce.

To add further evidence for the loopy nature (i.e. intrinsic discreteness) of the quantum bounce, a new avenue has been suggested in \cite{Chiou:2009hk} to consider a well-motivated extension of higher order holonomy corrections. At the level of heuristic effective dynamics, the investigation of \cite{Chiou:2009hk} reveals that the key features of the bouncing scenario are generic, regardless of the inclusion of higher order corrections. This strongly urges one to formulate the quantum theory of LQC with the higher order holonomy corrections and attest the observations of \cite{Chiou:2009hk} on a firmer ground. In this paper, by applying the techniques introduced in \cite{Ashtekar:2007em} to construct the physical Hilbert space and identify a complete family of Dirac observables, the quantum theory of LQC is rigorously formulated for any order $n$ of holonomy corrections (including the case of $n=\infty$). The detailed investigation shows that, for any arbitrary physical states (not only restricted to the states which are semiclassical at late times) and at any order $n$ of holonomy corrections (including $n=\infty$), the quantum bounce is generic and the expectation value of the matter density is bounded from above by an absolute upper bound, which equals the critical density obtained in the heuristic analysis of \cite{Chiou:2009hk}. By the same notion of ``singularity resolution'' suggested in \cite{Ashtekar:2007em}, the classical singularity is said to be resolved. With the extension of higher order holonomy corrections, we affirm to a broader extent the robustness of key features of LQC previously shown in \cite{Ashtekar:2007em}, the model of which is now viewed as a special case at the lowest order ($n=0$) of holonomy corrections.

Meanwhile, the higher order holonomy corrections might be understood as a result of using generic $SU(2)$ representations (labeled by spin $j$, not restricted to $j=1/2$) for the holonomies in the Hamiltonian constraint operator. To explore this possibility, we construct the Hamiltonian constraint operator in generic $j$ representations in a fashion slightly modified from that of \cite{Vandersloot:2005kh} and then investigate the linear sum of the Hamiltonian operators over generic $j$. The investigation is not conclusive but encouraging and reminiscent of the issues regarding large $j$ in some spin-foam models, calling for further research on the $j$ ambiguity from a new perspective.

This paper is organized as follows. After a brief overview on the motivations in \secref{sec:motivations}, the main text is divided into two major parts. In \secref{sec:PART I} as the first part, we study the Hamiltonian constraint operator in generic $j$ representations and their linear superposition. In \secref{sec:PART II} as the second part, the quantum theory of LQC with higher order holonomy corrections is formulated and used to investigate the physical dynamics. Finally, the conclusions are summarized and discussed in \secref{sec:discussion}.

\section{Motivations}\label{sec:motivations}
In the presence of a free massless scalar field $\phi(\vec{x},t)=\phi(t)$, the classical Hamiltonian constraint for the $k=0$ FRW model is given by
\be\label{eqn:classical C}
C=C_\grav+C_\matt
=-\frac{3N}{8\pi G\gamma^2}\,c^2\sqrt{\abs{p}}+\frac{Np_\phi^2}{2\,\abs{p}^{3/2}}
\ee
in terms of the Ashtekar variables $c$ and $p$, the conjugate momentum $p_\phi$ of $\phi$, the lapse function $N$, and the Barbero-Immirzi parameter $\gamma$.
At the heuristic level, to impose the loop quantum corrections of LQC, we simply take the prescription of ``holonomization'' to replace $c$ with
\be\label{eqn:holonomization}
c \longrightarrow \frac{\sin(\mubar c)}{\mubar}
\ee
by introducing a discreteness variable $\mubar$. The heuristic effective dynamics is then solved as if the dynamics was classical but governed by the new ``holonomized'' Hamiltonian, which reads as
\be\label{eqn:C mubar}
C_\mubar
=-\frac{3N}{8\pi G\gamma^2}\frac{\sin^2\mubar c}{\mubar^2}
\sqrt{\abs{p}}+\frac{Np_\phi^2}{2\,\abs{p}^{3/2}}.
\ee
The bouncing scenario can be easily obtained at the level of heuristic effective dynamics without invoking the sophisticated features of LQC \cite{Chiou:2007dn,Chiou:2007sp,Chiou:2007mg}, therefore calling into question whether the quantum bounce results from the very quantum nature of LQG.
In particular, with the ``improved'' scheme \eqnref{eqn:mu bar} imposed for $\mubar$, the modified Hamiltonian constraint $C_\mubar=0$ immediately sets an upper bound for the matter density:
\be
\rho_\phi:=\frac{p_\phi^2}{2\abs{p}^3}=\frac{3}{8\pi G\gamma^2\Delta}
\sin^2\mubar c\leq 3\rho_\mathrm{Pl},
\ee
where the Planckian density is defined as
\be\label{eqn:Planckian density}
\rho_\mathrm{Pl}:=(8\pi G \gamma^2\Delta)^{-1}.
\ee

Apparently, without going into the detailed construction of LQC at all, it is anticipated that the matter density is bounded from above and thus the quantum bounce occurs. One might then argue that the boundedness of $\rho_\phi$ has little to do with the fundamental structure of LQC but merely an \textit{ad hoc} consequence of the formal modification of \eqnref{eqn:holonomization}. In this sense, the quantum bounce seems not really ``loopy'' enough.

In regard to this issue, a well-motivated extension of higher order holonomy corrections has been suggested in \cite{Chiou:2009hk} and used to test whether the quantum bounce is essentially a consequence of the loopy nature (i.e. intrinsic discreteness) of LQC. By evoking the Taylor series:
\be
\sin^{-1}x=\sum_{k=0}^\infty \frac{(2k)!}{2^{2k}(k!)^2(2k+1)}\,x^{2k+1}
\ee
for $-1\leq x \leq 1$, the $n$th order holonomized connection variable is defined as
\be\label{eqn:holonomized c}
c_h^{(n)}:=\frac{1}{\mubar}\sum_{k=0}^{n} \frac{(2k)!}{2^{2k}(k!)^2(2k+1)}\,
{(\sin \mubar c)}^{2k+1},
\ee
which can be made arbitrarily close to $c$ (as $n\rightarrow\infty$) but remains a function of the holonomy $\sin \mubar c$ and the discreteness variable $\mubar$. To implement the underlying structure of LQC by replacing $c$ with holonomies, $c_h^{(n)}$ can be used as an improved version of \eqnref{eqn:holonomization}, which now reads as $c_h^{(n=0)}$. In place of \eqnref{eqn:C mubar}, the Hamiltonian constraint with holonomy corrections up to the $n$th order is then designated as
\be\label{eqn:C n}
C^{(n)}=C^{(n)}_\grav+C_\matt
:=-\frac{3N}{8\pi G\gamma^2}\,(c_h^{(n)})^2
\sqrt{\abs{p}}+\frac{Np_\phi^2}{2\,\abs{p}^{3/2}},
\ee
which supposedly reflects the intrinsic discreteness of LQC in a more elaborate fashion.

One might suspect that if we include corrections of all orders ($n=\infty$), the quantum theory of $C^{(n=\infty)}$ will lead to the same result as the WDW theory, as formally we have $c_h^{(\infty)}=c$. This should not be the case, because the elementary variables are still $\sin \mubar c$ and $p$, instead of $c$ and $p$, even in the limit $n\rightarrow\infty$ and, therefore, the striking difference between LQC and the WDW theory as emphasized in \cite{Ashtekar:2007em} should persist.

The analysis of \cite{Chiou:2009hk} at the level of heuristic effective dynamics indeed shows that the bouncing scenario is generic, irrespective of the order of corrections, and the matter density remains finite, bounded from above by an upper bound in the regime of the Planckian density, even if all orders of corrections are included. In addition to \cite{Ashtekar:2007em}, this observation adds further evidence that the intrinsic discreteness of LQC inherently accounts for the quantum bounce and thus LQC is fundamentally different from the WDW theory. It also encourages one to construct the quantum theory of LQC with the higher order holonomy corrections.

In \secref{sec:PART II}, we formulate the quantum theory of LQC associated with  the $n$th order Hamiltonian constraint:
\be\label{eqn:Cprime n}
C^{\prime(n)}=C^{\prime(n)}_\grav+C'_\matt
:=-\frac{3}{8\pi G\gamma^2}\,(c_h^{(n)})^2 p^2
+\frac{p_\phi^2}{2},
\ee
which is rescaled from \eqnref{eqn:C n} with $N=\abs{p}^{3/2}$. Not only can LQC based on \eqnref{eqn:Cprime n} be rigorously constructed but it also confirms the bouncing scenario to be generic, regardless of the order $n$, with the matter density bounded from above by an absolute upper bound, which equals the critical density obtained at the heuristic level in \cite{Chiou:2009hk}.

The higher order holonomy corrections corresponding to higher powers of $\sin\mubar c$ are most likely to be interpreted as the imprint of generic $j$ representations for holonomies in the Hamiltonian constraint operator. \secref{sec:PART I} is devoted to investigate this possibility. The aim is to show that, with proper regularization, the linear superposition of the Hamiltonian operators in generic $j$ representations might conspire to give the operator of $C^{{(n=\infty)}}_\grav$ or $C^{\prime{(n=\infty)}}_\grav$; i.e. given with appropriate coefficients $c_j$, we might have
\be\label{eqn:linear sum}
\sum_{j}c_j\,{}^{(j)}\hat{C}_\grav
=\hat{C}^{{(\infty)}}_\grav,
\ee
or
\be\label{eqn:linear sum prime}
\sum_{j}c_j\,{}^{(j)}\hat{C}'_\grav
=\hat{C}^{\prime{(\infty)}}_\grav,
\ee
where ${}^{(j)}\hat{C}'_\grav$ is the gravitational part of the (rescaled) Hamiltonian operator in the generic $j$ representation and $\hat{C}^{\prime{(\infty)}}_\grav$ is the operator corresponding to $C^{\prime(n=\infty)}_\grav$. This evokes the idea in \cite{Gaul:2000ba} that the extension of the Hamiltonian constraint operator to generic $j$ representations is motivated by the spacetime covariant formulation (such as spin-foam theory) of quantum gravity. In particular, the requirement of \emph{crossing symmetry} as manifested in the covariant formulation naturally leads one to consider the Hamiltonian operators in $j$, which add links of arbitrary colors of $j$. A linear combination of such operators could actually exist and define a crossing-symmetric quantization of the Hamiltonian constraint of general relativity. On the other hand, the general considerations in \cite{Vandersloot:2005kh} for LQC and in \cite{Perez:2005fn} for LQG both disfavor the Hamiltonian operator in generic $j$ representations other than $j=1/2$. The fact that the problem suggested in \cite{Vandersloot:2005kh,Perez:2005fn} could be avoided is commented on at the end of \secref{sec:bouncing scenario}.

\ \newline\noindent
\textbf{Remark:}

There are many other avenues of adding higher order holonomy corrections in the literature which are motivated differently and yield distinct dynamical behavior. In particular, the difference between the consideration here and that of \cite{Mielczarek:2008zz,Hrycyna:2008yu} is commented on in \cite{Chiou:2009hk}.

\section{Hamiltonian operators with higher order holonomy corrections}\label{sec:PART I}
This section presents the first half part of the paper. In \secref{sec:classical Hamiltonian}, we briefly review the notation and background knowledge for the classical theory (see \cite{Ashtekar:2006uz} for more details). In \secref{sec:Loop quantization in generic j}, following the treatment in \cite{Vandersloot:2005kh}, we construct the (rescaled) Hamiltonian operator of LQC in a generic $j$ representation with some simplifications. Finally, \secref{sec:linear superposition of Hamiltonian operators} investigates the linear sum of the Hamiltonian operators over generic $j$. Our goal is not to reach a definitive conclusion that a specific linear superposition of the Hamiltonian operators can give rise to the operator associated with $C^{\prime(\infty)}_\grav$ as suggested in \eqnref{eqn:linear sum prime}, but instead we try to justify this possibility and hopefully to inspire further research on the issues of $j$ ambiguity from this point of view.

\subsection{Classical Hamiltonian constraint}\label{sec:classical Hamiltonian}
In the standard Hamiltonian treatment of the cosmological model, the spacetime $M$ is foliated as $M=\Sigma\times \mathbb{R}$ with $\Sigma$ being the homogeneous spacelike slice. In the $k=0$ FRW model, because $\Sigma$ is noncompact, one has to introduce a finite-sized elementary cell $\mathcal{V}$ on $\Sigma$ and restricts all spatial integrations to $\mathcal{V}$. In the \emph{comoving} coordinates $\vec{x}=(x,y,z)$ in which the spacetime metric reads as $ds^2=g_{\mu\nu}dx^\mu dx^\nu=-dt^2+a(t)^2d\vec{x}^{\,2}$, $\mathcal{V}$ is chosen to be a cubic cell with \emph{fixed} coordinate length $L$ on each edge. With the symmetries of homogeneity and isotropy imposed, the gravitational phase space variables (the connections ${A_a}^i$ and the density weighted triads ${\tE^a}_i$) can be expressed as
\begin{subequations}
\ba
{A_a}^i&=&c\,V_o^{-\frac{1}{3}}\,\fcotriad{a}{i},\\
{\tE^a}_i&=&p\,\sqrt{\fq}\,V_o^{-\frac{2}{3}}\,\ftriad{a}{i}.
\ea
\end{subequations}
Here, $\ftriad{a}{i}$ and $\fcotriad{a}{i}$ are a set of \emph{fiducial} triads and cotriads ($\ftriad{a}{i}$ is the inverse of $\fcotriad{a}{i}$) that are left invariant by the action of the Killing fields of $\Sigma$ and adapted to be aligned with the edges of $\mathcal{V}$ (i.e. $\ftriad{a}{i}\propto\delta_i^a$ and $\fcotriad{a}{i}\propto\delta_a^i$). The fiducial 3-metric is given by $\fq_{ab}=\delta_{ij}\fcotriad{a}{i}\,\fcotriad{b}{j}$ and $V_o$ is the fiducial volume defined as the volume of $\mathcal{V}$ measured with respect to $\fq_{ab}$, i.e. $V_o:=\int_\mathcal{V}d^3x\sqrt{\fq}\equiv\int_0^L\int_0^L\int_0^L d^3x\sqrt{\fq}=L^3\sqrt{\fq}$ with $\fq:=\det{\fq_{ab}}$. Consequently, the symmetry-reduced phase space is described by the pair of $c$ and $p$, which satisfy the canonical relation:
\be\label{eqn:canonical rel of c and p}
\{c,p\}=\frac{8\pi G\gamma}{3}.
\ee
The absolute value of $p$ gives the \emph{physical} area of each rectangular surface of $\mathcal{V}$ while the sign of $p$ corresponds to the spatial orientation.\footnote{On the other hand, the physical interpretation of $c$ depends on the dynamics. For classical dynamics, we have $c=\gamma L \dot{a}(t)$, i.e. the change rate of the \emph{physical} length of the edges of $\mathcal{V}$ (times $\gamma$) with respect to the proper time $t$.}

In terms of $c$ and $p$, the gravitational part of the Hamiltonian constraint can be written as
\be\label{eqn:Cgrav}
C_\grav=-\frac{1}{16\pi G\gamma^2}\int_\mathcal{V} d^3x\, N e^{-1} {\epsilon_i}^{jk}F^i_{ab}{\tE^a}_j{\tE^b}_k
=-\frac{3N}{8\pi G\gamma^2}\,c^2\sqrt{\abs{p}}\,,
\ee
where
\be\label{eqn:e}
e:=|\det\tilde{E}|^{1/2}=L^{-3}\abs{p}^{3/2}
\ee
and $N$ is the lapse function.
Additionally, with the inclusion of a free massless scalar field, the total Hamiltonian constraint also contains the matter part given by
\be\label{eqn:Cmatt}
C_\matt=\int_\mathcal{V}d^3x\,N e \frac{\dot{\phi}^2}{2}
=\frac{N p_\phi^2}{2\,\abs{p}^{3/2}},
\ee
where $p_\phi:=\abs{p}^{3/2}\dot{\phi}$ is the conjugate momentum of the homogeneous field $\phi(\vec{x},t)=\phi(t)$ with
\be\label{eqn:canonical rel of phi and pphi}
\{\phi,p_\phi\}=1.
\ee
The classical dynamics is governed by the total Hamiltonian constraint $C=C_\grav+C_\matt$ as given in \eqnref{eqn:classical C}.

\subsection{Loop quantization in generic $j$ representations}
\label{sec:Loop quantization in generic j}
In the passage to quantum theory, the quantization procedure of LQC is more intricate than that of the WDW theory. This is because the kinematical Hilbert space of LQC is in the ``polymer representation'' for $p$, instead of the standard Schr\"{o}dinger representation. Consequently, the operator corresponding to $c$ does not exist and additionally it is not densely defined if the inverse volume function $\abs{p}^{-3/2}$ is na\"{i}vely quantized as the operator with eigenvalues equal  to the inverse of the volume eigenvalues.
Thus, to construct the Hamiltonian constraint operator, we have to express the classical constraint in terms of the triad variable $p$ and the holonomy $h_k^{(\mubar)}$, both of which have direct quantum analogs.

Following the close similarity of the regularization procedure \`{a} la Thiemann's trick used in the full theory \cite{Thiemann:2007zz}, we first recast the term involving the inverse triad $e^{-1}$ as
\be\label{eqn:EE}
e^{-1}{\epsilon_i}^{jk}{\tE^a}_j{\tE^b}_k
=\sum_l \frac{\fq^{1/3}}{2\pi G\gamma\,\mubar L}\,
\epsilon^{jkl}\,\ftriad{a}{j}\,\ftriad{b}{k}\,
\Tr\left(h_l^{(\mubar)}\{{(h_l^{(\mubar)})}^{-1},V\}\tau_i\right),
\ee
where
\be
h_l^{(\mubar)}:=\mathcal{P}\exp\int_0^{\mubar L} \tau_i {A_a}^i dx^a
=\exp(\mubar c\tau_l)
\ee
is the \emph{holonomy} along the edge of coordinate length $\mubar L$ aligned with the direction of $\ftriad{a}{l}\partial_a$, $V=\abs{p}^{3/2}$ is the physical volume of $\mathcal{V}$, and $\tau_i$ are the $SU(2)$ generators with $[\tau_i,\tau_j]={\epsilon_{ij}}^k\tau_k$ ($2i\tau_i=\sigma_i$ are the Pauli matrices in the standard convention). Note that this identity holds for any
choice of $\mubar$, even when it is allowed to be a function of $p$.

Next, invoking the standard techniques in gauge theories, Stokes's theorem allows us to express the field strength component $F^i_{ab}$ in terms of holonomies as
\be\label{eqn:Fiab}
F^i_{ab}\approx
-2\Tr\left[\left(h^{(\mubar)}_{\Box_{jk}}-1\right)\tau^i\right]
\frac{\fq^{-1/3}}{\mubar^2L^2}\,\fcotriad{a}{j}\,\fcotriad{b}{k},
\ee
where
\be
h^{(\mubar)}_{\Box_{jk}}:=h_j^{(\mubar)}h_k^{(\mubar)}
{(h_j^{(\mubar)})}^{-1}{(h_k^{(\mubar)})}^{-1}
\ee
is the holonomy around the boundary of the square $\Box_{jk}$, which is perpendicular to the direction of $\ftriad{a}{i}\partial_a$ and of coordinate length $\mubar L$ on each edge.
In the context of homogeneous models, we have $F^i_{ab}=\partial_{[a}{A_{b]}}^j+{\epsilon^i}_{jk}{A_a}^j{A_b}^k ={\epsilon^i}_{jk}{A_a}^j{A_b}^k$; thus, Stokes's theorem is not essential in \eqnref{eqn:Fiab} and we can simply express $\tau_j{A_a}^j\,\ftriad{a}{i}=\fq^{-1/6}\lim_{\mubar\rightarrow0} \big(h_i^{(2\mubar)}-(h_i^{(2\mubar)})^{-1}\big)/(4\mubar L)$, which leads to an alternative expression:
\be\label{eqn:Fiab 2}
F^i_{ab}\approx
-2\Tr\left([h_j^{(2\mubar)}-(h_j^{(2\mubar)})^{-1},
h_k^{(2\mubar)}-(h_k^{(2\mubar)})^{-1}]\tau^i\right)
\frac{\fq^{-1/3}}{(4\mubar L)^2}\,\fcotriad{a}{j}\,\fcotriad{b}{k},
\ee
where the extra factor $2$ in $h_i^{(2\mubar)}$ is adopted for later convenience as will be seen. In the continuous limit $\mubar\rightarrow0$, both \eqnref{eqn:Fiab} and \eqnref{eqn:Fiab 2} become exact; however, in the theory of LQC, this limit does not exist and we should keep $\mubar$ \emph{finite} as the ``regulator'' which reflects the ``fundamental discreteness'' of the full theory of LQG as an imprint on the reduced theory. To quantize $C_\mathrm{grav}$ as an operator, keeping $\mubar$ finite, we depart from \eqnref{eqn:EE} with either \eqnref{eqn:Fiab} or \eqnref{eqn:Fiab 2}. Moreover, when the sum of higher $j$ representations is considered in the quantization procedure, it turns out \eqnref{eqn:Fiab 2} is more controllable than \eqnref{eqn:Fiab} and thus is preferred.\footnote{The Hamiltonian operator based on \eqnref{eqn:Fiab 2} instead of \eqnref{eqn:Fiab} has been investigated in \cite{Yang:2009nj} in $j=1/2$ representation. It shows that the alternative quantization gives rise to the operator $\hat{C}_\mathrm{grav}=4\sin\frac{\mubar c}{2}\,\hat{A}\sin\frac{\mubar c}{2}$, the dynamics of which is only slightly different from that of the operator $\hat{C}_\mathrm{grav}=\sin\mubar c\,\hat{A}\sin\mubar c$ obtained in the standard approach \eqnref{eqn:Fiab} with $\hat{A}$ being the operator quantized from \eqnref{eqn:EE}. In fact, if an extra factor $2$ in $h_i^{(2\mubar)}$ is also adopted as in \eqnref{eqn:Fiab 2}, the former is identical to the latter.}

To obtain the quantum theory, the holonomies $h_i^{(\mubar)}$ are promoted to ${}^{(j)}\hat{h}_i^{(\mubar)}$ in the $j$ representation, $V$ to $\hat{V}$, and the Poisson bracket to a commutator. In the $j$ representation of the $SU(2)$ group, the Lie algebra generators $\tau_i$ are represented as $(2j+1)\times(2j+1)$ matrices ${}^{(j)}\tau_i$, which satisfy
\be
\Tr\left({}^{(j)}\tau_i{}^{(j)}\tau_j\right)=-\frac{1}{3}j(j+1)(2j+1)\,\delta_{ij}.
\ee
By \eqnref{eqn:EE} and \eqnref{eqn:Fiab}, the classical Hamiltonian \eqnref{eqn:Cgrav} with $N=1$ then leads to the Hamiltonian operator in the generic $j$ representation:
\ba\label{eqn:Cgrav op}
{}^{(j)}\hat{C}_\grav&=&\frac{3i}{(8\pi G)^2\hbar\gamma^3j(j+1)(2j+1)\mubar^3}\nn\\
&&\times\sum_{ijk}\epsilon^{ijk}\,
\Tr\left({}^{(j)}\hat{h}_i^{(\mubar)} {}^{(j)}\hat{h}_j^{(\mubar)}
{{}^{(j)}\hat{h}_i^{(\mubar)}}^{-1} {{}^{(j)}\hat{h}_j^{(\mubar)}}^{-1}
{}^{(j)}\hat{h}_k^{(\mubar)}[\,{{}^{(j)}\hat{h}_k^{(\mubar)}}^{-1},\hat{V}\,]\right).
\ea
The Hamiltonian operator in generic $j$ representations has been studied in depth in \cite{Vandersloot:2005kh}. The matrix elements of holonomies in the $j$ representation are given by
\begin{subequations}\label{eqn:elements of h}
\ba
\left({}^{(j)}\hat{h}_1^{(\mubar)}\right)_{mn}
&=&T_{mn}\sum_{s=\abs{m-n},\abs{m-n}+2,\dots}^{2j-\abs{m+n}}
\frac{(-i)^s}{Y_{mns}}\,\widehat{\cos\mubar c/2}^{2j-s}\widehat{\sin\mubar c/2}^s,\\
\left({}^{(j)}\hat{h}_2^{(\mubar)}\right)_{mn}
&=&T_{mn}\sum_{s=\abs{m-n},\abs{m-n}+2,\cdots}^{2j-\abs{m+n}}
\frac{(-i)^{n-m+s}}{Y_{mns}}\,\widehat{\cos\mubar c/2}^{2j-s}\widehat{\sin\mubar c/2}^s,\\
\left({}^{(j)}\hat{h}_3^{(\mubar)}\right)_{mn}
&=&\widehat{e^{im\mubar c}}\,\delta_{mn},
\ea
\end{subequations}
where $m,n\in\{-j,-j+1,\dots,j-1,j\}$, the index $s$ in the sum increases by an increment of 2, and $T_{mn}$ and $Y_{mns}$ are constants given by
\begin{subequations}
\ba
T_{mn}&=&\sqrt{(j+m)!(j-m)!(j+n)!(j-n)!}\,,\\
Y_{mns}&=&\left(j+\frac{1}{2}(m+n-s)\right)!\left(j-\frac{1}{2}(m+n+s)\right)!\nn\\
&&\times\left(\frac{1}{2}(m-n+s)\right)!\left(\frac{1}{2}(n-m+s)\right)!.
\ea
\end{subequations}
The resulting Hamiltonian operator reads as
\ba\label{eqn:jCgrav}
{}^{(j)}\hat{C}_\grav&=&\frac{-9i}{(8\pi G)^2\hbar\gamma^3j(j+1)(2j+1)\mubar^3}\nn\\
&&\times
\sum_{m=-j}^{j}\sum_{s'=1,2,\cdots}^{4j-1}{}^{(j)}\!Z_m^{2s'}
\widehat{\cos\mubar c/2}^{8j-2s'}\widehat{\sin\mubar c/2}^{2s'}
\widehat{e^{im\mubar c}}\,\hat{V}\widehat{e^{-im\mubar c}},
\ea
where the constants ${}^{(j)}\!Z_m^s$ can be expressed in terms of $T_{mn}$ and $Y_{mns}$ (see \cite{Vandersloot:2005kh}).

Similarly, for the matter part of the Hamiltonian constraint, to deal with the inverse volume $V^{-1}=\abs{p}^{-3/2}$ in \eqnref{eqn:Cmatt}, Thiemann's trick is used again to define the inverse volume operator as
\be\label{eqn:jV-1}
{}^{(j)}\widehat{V^{-1}}=
\left[-\frac{3i\,\ell^{-1}}{8\pi G\hbar\gamma j(j+1)(2j+1)\mubar}
\sum_i\Tr\left({}^{(j)}\tau_i
{}^{(j)}\hat{h}_i^{(\mubar)}[\,{{}^{(j)}\hat{h}_i^{(\mubar)}}^{-1},
\hat{V}^\frac{2\ell}{3}]\right)
\right]^\frac{3}{2(1-\ell)}.
\ee
Here arise two ambiguities, labeled by a half-integer $j$ for different choices of $j$ representations and a real number $\ell$ ($0<\ell<1$) \cite{Bojowald:2002ny}. The general considerations in \cite{Vandersloot:2005kh,Perez:2005fn} urge one to set $j=1/2$.\footnote{Since we will consider the linear sum of the Hamiltonian operators over all values of $j$, it is not clear whether one should still stick with $j=1/2$ for the inver volume operator. This problem will not bother us, as we will rescale the Hamiltonian with an appropriate $N$ before quantization to avoid the Thiemann's trick. See the remark in the end of \secref{sec:linear superposition of Hamiltonian operators} for more comments.} For $\ell$, a general criterion is not available and $\ell=1/2$ and $\ell=3/4$ are most used in the literature.

Unfortunately, the formula \eqnref{eqn:jCgrav} does not yield the desirable expression which would readily match \eqnref{eqn:linear sum prime} when the operators ${}^{(j)}\hat{C}_\grav$ of generic $j$ are properly summed. In addition, the inverse volume operator in \eqnref{eqn:jV-1} gives rise to further complications. To achieve our goal of having \eqnref{eqn:linear sum}, we make two simplifications: First, we neglect the problems involving the inverse triad by choosing an appropriate lapse function $N$ before quantization and thus obviate Thiemann's trick; second, instead of \eqnref{eqn:Fiab}, the alternative quantization scheme \eqnref{eqn:Fiab 2} is adopted for the field strength component.

In the full theory of LQG, it is not possible to get rid of the involvement of the inverse triad by gauge fixing $N$ to eliminate $e^{-1}$, because $e$ is a weight-one scalar density while $N$ is a scalar with respect to diffeomorphisms on $\Sigma$. However, in the context of the $k=0$ FRW model, thanks to homogeneity, we have \eqnref{eqn:e} and thus the difficulty due to the inverse triad can be readily avoided by choosing $N=\abs{p}^{3/2}$.\footnote{Note that the choice of $N=\abs{p}^{3/2}$ gives a scalar since $\abs{p}$ is the \emph{physical} area of the faces of $\mathcal{V}$ and independent of the coordinates; on the other hand, $e=L^{-3}\abs{p}^{3/2}$ yields a weight-one scalar density because of the coordinate-dependent factor $L^{-3}$.} With this choice of $N$, in parallel to \eqnref{eqn:Cgrav} and \eqnref{eqn:Cmatt}, the rescaled Hamiltonian $C'=C'_\grav+C'_\matt$ reads as
\be\label{eqn:Cgrav prime}
C'_\grav=-\frac{1}{16\pi G\gamma^2}\int_\mathcal{V} d^3x\,L^3 {\epsilon_i}^{jk}F^i_{ab}{\tE^a}_j{\tE^b}_k
=-\frac{3}{8\pi G\gamma^2}\,c^2\abs{p}^2
\ee
and
\be\label{eqn:Cmatt prime}
C'_\matt=\int_\mathcal{V}d^3x\frac{\abs{p}^3}{L^3} \frac{\dot{\phi}}{2}
=\frac{p_\phi^2}{2},
\ee
as given in \eqnref{eqn:Cprime n}.
With this gauge fixing imposed \emph{before} quantization, the quantization strategy is much easier and does not need Thiemann's trick anymore.

Furthermore, the second simplification is to adopt \eqnref{eqn:Fiab 2} instead of \eqnref{eqn:Fiab}. In parallel to \eqnref{eqn:Cgrav op}, taking \eqnref{eqn:Fiab 2} into \eqnref{eqn:Cgrav prime} yields the (rescaled) Hamiltonian operator in the generic $j$ representation given as
\begin{subequations}\label{eqn:Cgrav prime op}
\ba
\label{eqn:Cgrav prime op part a}
{}^{(j)}\hat{C}'_\grav&=&\frac{3}{8\pi G\gamma^2j(j+1)(2j+1)(4\mubar)^2}\nn\\
&&\times\sum_{ijk}\epsilon^{ijk}\,
\Tr\left([\,{}^{(j)}\hat{h}_j^{(2\mubar)}-({}^{(j)}\hat{h}_j^{(2\mubar)})^{-1},
{}^{(j)}\hat{h}_k^{(2\mubar)}-({}^{(j)}\hat{h}_k^{(2\mubar)})^{-1}]
{}^{(j)}\tau_i\right)
\abs{\hat{p}}^{2}\quad\\
\label{eqn:Cgrav prime op part b}
&=&
\frac{18}{8\pi G\gamma^2j(j+1)(2j+1)(4\mubar)^2}\nn\\
&&\times
\Tr\left[\left({}^{(j)}\hat{h}_1^{(2\mubar)}-({}^{(j)}\hat{h}_1^{(2\mubar)})^{-1}\right)
\left({}^{(j)}\hat{h}_2^{(2\mubar)}-({}^{(j)}\hat{h}_2^{(2\mubar)})^{-1}\right)
{}^{(j)}\tau_3\right]
\abs{\hat{p}}^{2},
\ea
\end{subequations}
where the fact that the operator is $SU(2)$ invariant has been used to arrive at \eqnref{eqn:Cgrav prime op part b}.
Using \eqnref{eqn:elements of h}, we have
\begin{subequations}\label{eqn:expression for Tr}
\ba
&&\Tr\left[\left({}^{(j)}\hat{h}_1^{(2\mubar)}-({}^{(j)}\hat{h}_1^{(2\mubar)})^{-1}\right)
\left({}^{(j)}\hat{h}_2^{(2\mubar)}-({}^{(j)}\hat{h}_2^{(2\mubar)})^{-1}\right)
{}^{(j)}\tau_3\right]\nn\\
\label{eqn:expression for Tr part a}
&=&
\sum_{m,n=-j}^{j} (i m)
\sum_{s_1,s_2=\abs{m-n},\abs{m-n}+2,\cdots}^{2j-\abs{m+n}}
\frac{T_{mn}T_{nm}(-i)^{m-n+s_1+s_2}}
{Y_{mns_1}Y_{nms_2}}\nn\\
&&\quad\times\left[1-(-1)^{s_1}\right]\left[1-(-1)^{s_2}\right]
\widehat{\cos\mubar c}^{\,4j-s_1-s_2}\,
\widehat{\sin\mubar c}^{\,s_1+s_2}\\
\label{eqn:expression for Tr part b}
&=:&\sum_{s=2,4,\cdots}^{4j}{}^{(j)}\!W_s\,
\widehat{\cos\mubar c}^{\,4j-s}\,
\widehat{\sin\mubar c}^{\,s}
=:\sum_{k=1,2,\cdots}^{2j}{}^{(j)}\!a_k\,\widehat{\sin\mubar c}^{\,2k},
\ea
\end{subequations}
where the index $s$ is simply $s=s_1+s_2$ and we have used the fact that the summand in the sum is nonvanishing only if $s_1$ and $s_2$ are both odd to arrive at \eqnref{eqn:expression for Tr part b}. This leads to
\be\label{eqn:Cgrav prime op 2}
{}^{(j)}\hat{C}'_\grav=\frac{18}{8\pi G\gamma^2j(j+1)(2j+1)(4\mubar)^2}
\left(\sum_{k=1,2,\cdots}^{2j}{}^{(j)}\!a_k\,\widehat{\sin\mubar c}^{\,2k}\right)
\abs{\hat{p}}^{2}.
\ee
Explicitly, we list ${}^{(j)}\hat{C}'_\grav$ for the leading values of $j$:
\begin{subequations}
\ba
{}^{(1/2)}\hat{C}'_\grav&=&-\frac{3}{8\pi G\gamma^2\mubar^2}\,
\widehat{\sin\mubar c}^{\,2}\abs{\hat{p}}^{2},\\
{}^{(1)}\hat{C}'_\grav&=&-\frac{3}{8\pi G\gamma^2\mubar^2}\,
\left(\widehat{\sin\mubar c}^{\,2}-\widehat{\sin\mubar c}^{\,4}\right)\abs{\hat{p}}^{2},\\
{}^{(3/2)}\hat{C}'_\grav&=&-\frac{3}{8\pi G\gamma^2\mubar^2}\,
\left(\widehat{\sin\mubar c}^{\,2}
-\frac{12}{5}\,\widehat{\sin\mubar c}^{\,4}
+\frac{6}{5}\,\widehat{\sin\mubar c}^{\,6}\right)
\abs{\hat{p}}^{2},\\
{}^{(2)}\hat{C}'_\grav&=&-\frac{3}{8\pi G\gamma^2\mubar^2}\,
\left(\widehat{\sin\mubar c}^{\,2}
-\frac{21}{5}\,\widehat{\sin\mubar c}^{\,4}
+\frac{24}{5}\,\widehat{\sin\mubar c}^{\,6}
-\frac{8}{5}\,\widehat{\sin\mubar c}^{\,8}\right)
\abs{\hat{p}}^{2}.
\ea
\end{subequations}
Note that the leading coefficient is given by ${}^{(j)}\!a_{k=1}=8j(j+1)(2j+1)/3$, which is in accord with the fact that, for any generic $j$, ${}^{(j)}\hat{C}'_\grav$ returns to the classical counterpart in the formal limit $\mubar\rightarrow0$.

\ \newline\noindent
\textbf{Remark:}

It has been hotly debated whether it is legitimate to rescale the Hamiltonian by the lapse function $N$ before quantization. This question is in fact asked in two respects: (i) for the theory of LQC derived from LQG; and (ii) within the confines of LQC. While the quest for a systematic formulation to derive LQC from LQG remains wide open, it is logically viable to obviate the issues of the inverse triad by gauge fixing $N$ within the confines of LQC. One might still argue that, without the corrections by Thiemann's trick, the resulting LQC would no longer have the appealing feature of singularity resolution, as it has been shown that the state in the kinematical Hilbert space associated with the classical singularity is decoupled in the evolution equation as a result of the loop corrections on the inverse triad \cite{Bojowald:2001vw}. However, decoupling of the singular state is neither a necessary nor a sufficient condition for singularity resolution. To provide a satisfactory notion of ``singularity resolution'', one should have available the physical Hilbert space and a complete family of Dirac observables. This issue is discussed in depth in \cite{Ashtekar:2007em}, and therein an applicable and satisfactory notion is suggested and used to show that the singularity is resolved and the corrections on the inverse triad are not essential. We will adopt the same notion of singularity resolution and reach the same conclusion for the quantum theory with higher order holonomy corrections in \secref{sec:singularity resolution}.

\subsection{Linear superposition of the Hamiltonian operators in generic $j$}
\label{sec:linear superposition of Hamiltonian operators}

As shown in \eqnref{eqn:Cgrav prime op 2}, ${}^{(j)}\hat{C}'_\grav$ is given as the sum of even powers of ${\sin \mubar c}$ up to $\sin^{4j} \mubar c$. As the right-hand side of \eqnref{eqn:linear sum prime} also consists of even powers of ${\sin \mubar c}$, the coefficients $c_j$ can be obtained through the truncated version of \eqnref{eqn:linear sum prime} as follows. For a given $n$, we can uniquely determine the set of constants $c_j^{(n)}$ such that
\be\label{eqn:linear superposition}
\sum_{j=1/2,1,\cdots}^{n/2} c_j^{(n)}\, {}^{(j)}\hat{C}'_\grav
=-\frac{3}{8\pi G\gamma^2\mubar^2}
\sum_{k=1}^{n}
b_k\,\widehat{\sin \mubar c}^{\,2k}
\abs{\hat{p}}^{2}
=:\hat{\tilde{C}}^{\prime(n)}_\grav,
\ee
where $b_k$ are the coefficients of the Taylor series:
\be
\left(\sin^{-1}x\right)^2=\sum_{k=1}^\infty b_k\, x^{2k}
\ee
$-1\leq x \leq 1$
and the operator $\hat{\tilde{C}}^{\prime(n)}_\grav$ is defined as the truncated version of $\hat{C}^{\prime{(\infty)}}_\grav$ (up to a factor ordering). That is, for any arbitrary $n$, we can always find a linear superposition of ${}^{(j)}\hat{C}'_\grav$ to match up the Hamiltonian operator $\hat{C}^{\prime{(\infty)}}_\grav$ with the terms of higher than $2n$ powers of $\sin\mubar c$ all removed.\footnote{Alternatively, instead of $\hat{\tilde{C}}^{\prime(n)}_\grav$, we can use $\hat{C}^{\prime{(n)}}_\grav$ as a truncated form of $\hat{C}^{\prime{(\infty)}}_\grav$ and replace \eqnref{eqn:linear superposition} with
\be
\sum_{j=1/2,1,\cdots}^{n+1/2} c_j^{(n)}\, {}^{(j)}\hat{C}'_\grav
=\hat{C}^{\prime{(n)}}_\grav.\nn
\ee
However, $\hat{C}^{\prime{(n)}}_\grav$, the operator of $C^{\prime{(n)}}_\grav$ as defined in \eqnref{eqn:Cprime n}, does not precisely strip off $\hat{C}^{\prime{(\infty)}}_\grav$ the terms of higher than $4n+2$ powers of $\sin\mubar c$. At the limit $n\rightarrow\infty$, the alternative treatment nevertheless yields the similar profile of asymptotic behavior for $c_j^{(n)}$; the computer computation in this approach indeed suggests $c_j^{(n)}\approx 2c_j\, n\, e^{n+1/2}$ in parallel to \eqnref{eqn:limiting cj} for very large $n$, but the computation is far less efficient as the convergence is much slower.}

\begin{figure}
\begin{picture}(470,230)(0,0)

\put(50,0)
{
\scalebox{1.5}{\includegraphics{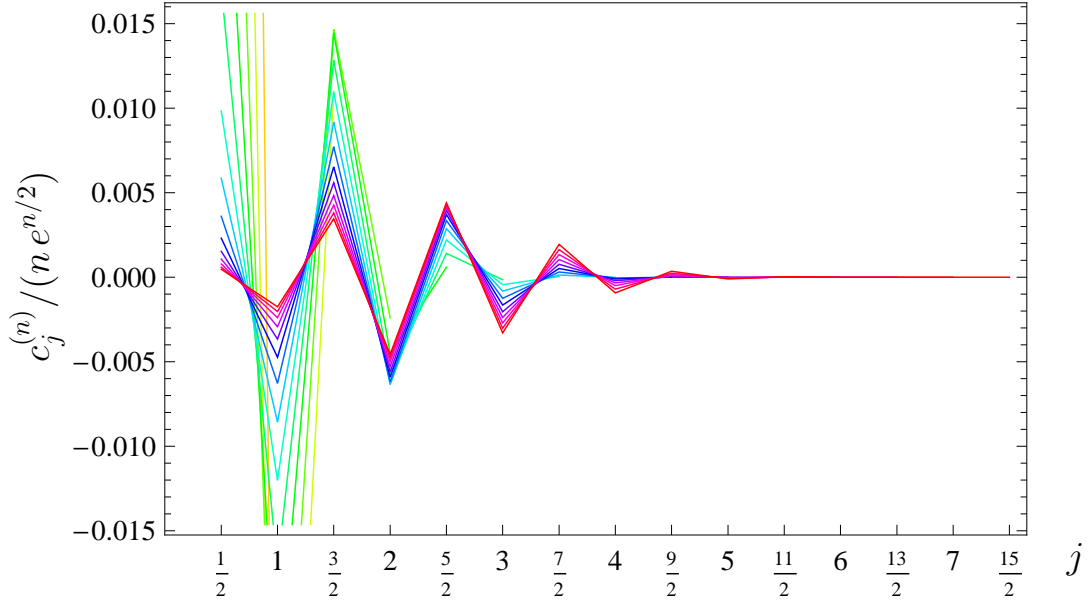}}
}

\put(430,12){\large $j$}
\put(30,90){\rotatebox{90}{\large $c_j^{(n)}/(n\,e^{n/2})$}}

\end{picture}
\caption{Coefficients $c_j^{(n)}$ scaled by $(n\,e^{n/2})^{-1}$ for $n=2,3,\dots,15$. ($c_j^{(n=1)}$ is a single point at $j=1/2$, which is out of the plot.) The scaled coefficients $c_j^{(n)}/(n\,e^{n/2})$ tend to converge to a set of constants $c_j$ as $n$ increases.}\label{fig:coefficients}
\end{figure}

Equation \eqnref{eqn:expression for Tr} gives a convenient algorithm to compute the coefficients ${}^{(j)}\!a_k$, which are then substituted to \eqnref{eqn:linear superposition} to compute $c_j^{(n)}$. The computation by computer shows that, undesirably, $c_j^{(n)}$ do not converge to a set of constants but rather diverge rapidly in magnitude as $n$ increases. Interestingly, a closer examination on the computed data suggests that the growth of $c_j^{(n)}$ yields a peculiar profile of asymptotic behavior:
\be\label{eqn:limiting cj}
c_j^{(n)}\approx c_j\, n\, e^{n/2},
\qquad
c_j: \text{ constants}
\ee
when $n$ is large enough, meaning that the \emph{relative} weights of $c_j^{(n)}$ nevertheless tend to be constant. As shown in \figref{fig:coefficients}, the coefficients $c_j^{(n)}$ rescaled by $(n\, e^{n/2})^{-1}$ tend to converge to a set of constants $c_j$ and the relative weights $c_j$ are appreciable only for moderate values of $j$ ($j\alt5$) but diminish rapidly for large $j$.\footnote{Solving \eqnref{eqn:linear superposition}, we have to deal with the linear equation $\mathbf{A}\vec{x}=\vec{b}$ with $\mathbf{A}$ being a $n\times n$ matrix. When $n$ is large, $\mathbf{A}$ contains both very big and very small entries. As a consequence, the double precision used in our computer program loses necessary accuracy and results in noticeable round-off error. To avoid technical difficulty, the numerical computation is done only up to $n=15$. Because of this limitation, we can only know the rough profile of $c_j$. That is, the asymptotic behavior in \eqnref{eqn:limiting cj} is only an outline suggested by the limited data of numerical computation and the literal expression should not be taken too seriously. Other formulae such as $c_j^{(n)}\approx c_j\, n^\alpha e^{n/2}$ with $\alpha\approx 1$ are also possible.} This asymptotic behavior suggests that, comparatively, only the contributions of moderate $j$ are manifest.

It is speculated that, if a regularization is imposed to suppress the very high $j$ contributions, $c_j^{(n)}$ can be tamed to converge to constants. More explicitly, a simple example is to regularize the matching condition \eqnref{eqn:linear superposition} as
\be
\sum_{j=1/2,1,\cdots}^{n/2} e^{-\delta f(j)}c_j^{(n)}\, {}^{(j)}\hat{C}'_\grav
=-\frac{3}{8\pi G\gamma^2\mubar^2}
\sum_{k=1}^{n}
e^{-\delta f(k)}\, b_k\, \widehat{\sin \mubar c}^{\,2k}
\abs{\hat{p}}^{2}
\ee
by introducing a monotonically increasing function $f(j)$ and a minuscule regulating parameter $\delta$ giving rise to a cut off for high $j$. With this regularization, $c_j^{(n)}$ are expected to converge to a set of constants.

The peculiar feature of divergence of $c_j^{(n)}$ associated with large $j$ is reminiscent of the infrared divergence encountered in many spin-foam models (see \cite{Ponzano-Regge,Freidel:2002dw} for the Ponzano-Regge model, \cite{Baez:1999sr} for the 4D BF model, and \cite{Oriti:2003wf} for a general review on spin-foam models). It is an infrared divergence, since it regards large $j$, namely, large lengths or areas, on the faces of spin foams. A renormalization procedure has been developed to divide away the divergence in the Ponzano-Regge model \cite{Ponzano-Regge,Freidel:2002dw}; the same technique might be applied here to mitigate the undesired divergence (see the remark below for more comments). Another appealing way to get rid of the infrared divergence of the spin-foam models is to replace the representation theory of the group ($SU(2)$ for the Ponzano-Regge model and $SO(4)$ for the 4D BF model) with that of the quantum group ($SU(2)_q$ and $SO(4)_q$, respectively, with $q$ chosen to be a root of unity). This leads to the Turaev-Viro model \cite{Turaev:1992hq} and the Crane-Yetter model \cite{Crane:1993if}, respectively. On both models, it can be argued that the quantum deformation of the group simply corresponds to the addition of a cosmological constant in the classical action, the value of which is given proportionally by $1/q$ (see \cite{Noui:2002ag} for this subject). It is tantalizing to speculate that the quantum deformation could also be used to remove the divergence of $c_j^{(n)}$; if this is indeed the case, the regulating parameter $\delta$ can be interpreted as a consequence of a small but nonzero cosmological constant.

The fact that the linear superposition of ${}^{(j)}\hat{C}'_\grav$, if suitably regulated, can yield the Hamiltonian operator $\hat{C}^{\prime(\infty)}_\grav$ bolsters the idea that it might be more natural to take into account all $j$ representations when the Hamiltonian constraint is quantized, although we have no theory yet to derive or even suggest the values of $c_j$ from the first principle. The resemblance between the divergence of $c_j^{(n)}$ and the infrared divergence in spin-foam models adds one more rationale to suggest that the inclusion of higher order holonomy corrections in the Hamiltonian operator may come out naturally from the spin-foam models (see also \cite{Gaul:2000ba}), although the link between LQC and the spin-foam formalism is far from clear. We leave these issues to future research and in the second half part of this paper we will simply take \eqnref{eqn:Cprime n} as our departing point to construct the corresponding quantum theory of LQC and study its dynamics.

\ \newline\noindent
\textbf{Remark:}

In the full theory of LQG, the classical Hamiltonian constraint has to be regulated before it is quantized. The procedure of the general classical regularization is described in \cite{Ashtekar:2004eh}. Assign a partition of $\Sigma$ into cells $\Box$ of possibly arbitrary shapes such that every cell $\Box$ has linear dimension smaller than $\epsilon$ in coordinates. For every cell $\Box$, we define edges $s_J$ and loops $\beta_{IJ}$ (lying on the surface spanned by $s_I$ and $s_J$). Finally, fix an arbitrary chosen representation $j$ of $SU(2)$. The entire structure is denoted as $R_\epsilon$ and called a \emph{permissible classical regulator} if
\be\label{eqn:regulator 1}
\lim_{\epsilon\rightarrow0}C_{R_\epsilon}^\mathrm{Eucl}(A,E)=
C^\mathrm{Eucl}(A,E),
\ee
where
\ba
\label{eqn:regulator 2}
C_{R_\epsilon}^\mathrm{Eucl}&=&\sum_\Box C_{R_\epsilon\Box}^\mathrm{Eucl}\\
\label{eqn:regulator 3}
C_{R_\epsilon\Box}^\mathrm{Eucl}&=&
\frac{N(v_\Box)}{(8\pi G)^2\gamma^{3/2}}
\sum_{IJK}C^{IJK}\Tr\left[
\left({}^{(j)}h_{\beta_{IJ}}-{}^{(j)}h_{\beta_{IJ}}^{-1}\right)
{}^{(j)}h_{s_K}^{-1}\{{}^{(j)}h_{s_K},V\}
\right]
\ea
with fixed constants $C_{IJK}$ and if a similar condition also holds for the Lorentzian part. If we go one step further to generalize this procedure, we could remove the restriction to a fixed representation of $SU(2)$ and thus include all generic $j$ representations. The notion of \emph{permissibility} of the classical regulator will be slightly changed as \eqnref{eqn:regulator 3} is modified to
\be\label{eqn:regulator 4}
C_{R_\epsilon\Box}^\mathrm{Eucl}=
\frac{N(v_\Box)}{(8\pi G)^2\gamma^{3/2}}
\sum_{j,IJK}{}^{(j)}C^{IJK}\Tr\left[
\left({}^{(j)}h_{\beta_{IJ}}-{}^{(j)}h_{\beta_{IJ}}^{-1}\right)
{}^{(j)}h_{s_K}^{-1}\{{}^{(j)}h_{s_K},V\}
\right]
\ee
with ${}^{(j)}C_{IJK}$ being constants.

In the context of LQC, the na\"{\i}ve analog to \eqnref{eqn:regulator 1} would be
\be
\lim_{\mubar\rightarrow0}{}^{(j)}C'_\grav=-\frac{3}{8\pi G\gamma^2}\,c^2\abs{p}^2
\ee
for a fixed $j$. However, the remarkable feature of LQG that the dependence of the regulating parameter $\epsilon$ disappears is no longer the case in LQC, because the diffeomorphism invariance is explicitly broken with homogeneity imposed. As a consequence, we should keep $\mubar$ finite by hand to impose the fundamental discreteness of LQG. Nevertheless, if we consider all generic $j$ representations rather than choose a fixed one, the notion of permissibility can be restored and the analogous criterion is given by
\be\label{eqn:regulator 1 LQC}
\lim_{n\rightarrow\infty}\tilde{C}^{\prime(n)}_\grav=-\frac{3}{8\pi G\gamma^2} \,c^2\abs{p}^2
\ee
with
\be\label{eqn:regulator 2 LQC}
\tilde{C}^{\prime(n)}_\grav=\sum_j c_j^{(n)}\, {}^{(j)}C'_\grav.
\ee
In a sense, the role of the regulating parameter $\epsilon$ in LQG has been traded for $n$ in LQC to achieve permissibility and, as a trade-off, the aforementioned divergence regarding large $j$ implies that the dependence of $n$ cannot be trivially removed. From this perspective, this divergence is analogous to that in the Ponzano-Regge spin-foam model, in which the finite triangulation (analogous to keeping $\mubar$ finite) breaks the continuous diffeomorphism invariance and a renormalization procedure taking into account the translation invariance has been developed to get rid of the divergence \cite{Freidel:2002dw}. A regularization scheme for the case of LQC might also have the same rationale.

\ \newline\noindent
\textbf{Remark:}

The virtue of adopting \eqnref{eqn:Fiab 2}, instead of \eqnref{eqn:Fiab}, and avoiding Thiemann's trick (by scaling the Hamiltonian with an appropriate $N$) is that the resulting Hamiltonian operator for a given $j$ involves only even powers of $\sin\mubar c$ as shown in \eqnref{eqn:Cgrav prime op 2}, giving a systematic algorithm to uniquely determine the coefficients $c_j^{(n)}$ via \eqnref{eqn:linear superposition}. Had we used \eqnref{eqn:Fiab} and/or taken into account Thiemann's trick, the resulting operator for a given $j$ would be more involved (see \cite{Vandersloot:2005kh} for the explicit expression of ${}^{(j)}\hat{C}_\grav$) and the matching condition in parallel to \eqnref{eqn:linear superposition} would be much more complicated. This does not necessarily impede the linear superposition of ${}^{(j)}\hat{C}_\grav$ to yield $\hat{C}^{{(\infty)}}_\grav$ as in \eqnref{eqn:linear sum}, but to show the possibility is much more difficult than in the alternative quantization scheme we have adopted. Furthermore, if Thiemann's trick is used to recast the term involving the inverse triad (although not strictly necessary as remarked at the end of \secref{sec:Loop quantization in generic j}), correspondingly, for the matter part, the inverse volume operator should also be defined \`{a} la Thiemann's trick as shown in \eqnref{eqn:jV-1}, which gives rise to both $j$ and $\ell$ ambiguities. As well as the gravitational part, the matter part of the Hamiltonian operator shall include all contributions of $j$. However, unlike the gravitational part, there is no clear sense to posit a linear combination of ${}^{(j)}\widehat{V^{-1}}$ as the most natural choice.

\section{Loop quantum cosmology with higher holonomy corrections}\label{sec:PART II}
This section presents the second half part of the paper. As the correspondence between higher order holonomy corrections and higher $j$ representations of the Hamiltonian operator has been explored in \secref{sec:PART I}, for the second part, we take \eqnref{eqn:Cprime n} for granted. Its quantum theory of LQC will be formulated and studied by applying the techniques introduced in \cite{Ashtekar:2007em}. In \secref{sec:physical Hilbert space}, we construct the physical Hilbert space and identify the complete set of Dirac observables. In \secref{sec:b and x representations}, the formulation is recast in the $b$ and $x$ representations to facilitate further calculations. In \secref{sec:singularity resolution}, we adopt the notion of singularity resolution suggested in \cite{Ashtekar:2007em} and prove that, for any arbitrary physical states and at any order $n$, the classical singularity is resolved as the matter density remains bounded from above by an absolute upper bound, which equals the critical density obtained in the heuristic analysis of \cite{Chiou:2009hk}. Finally, in \secref{sec:bouncing scenario}, we elaborate on the bouncing scenario for the case of $n=\infty$ (all orders of holonomy corrections included) and study the evolution of coherent (semiclassical) states.

\subsection{The full Hamiltonian constraint and the physical Hilbert space}
\label{sec:physical Hilbert space}
To impose the discreteness parameter $\mubar$, following the ``improved'' dynamics suggested by \cite{Ashtekar:2006wn}, the discreteness parameter $\mubar$ is designated as
\be\label{eqn:mu bar}
\mubar=\sqrt{\frac{\Delta}{\abs{p}}}\equiv\frac{\lambda}{\sqrt{\abs{p}}},
\ee
where $\lambda^2\equiv\Delta$ is the area gap in the full theory of LQG and $\Delta=2\sqrt{3}\pi\gamma\Pl^2$ for the standard choice (but other choices are also possible) with $\Pl:=\sqrt{G\hbar}$ being the Planck length. With this choice, it is convenient to introduce the new canonical variables:
\begin{subequations}
\ba
\nu&:=&\left(\sgn\,p\right)\frac{\abs{p}^{3/2}}{2\pi\gamma\Pl^2},\\
b&:=&\frac{c}{\sqrt{\abs{p}}},
\ea
\end{subequations}
which, by \eqnref{eqn:canonical rel of c and p}, satisfy the canonical relation
\be
\{b,\nu\}=2\hbar^{-1}.
\ee
Also note that $\mubar c=\lambda b$.

To construct the kinematical Hilbert space $\hil_\kin$ of LQC, it is convenient to use the $\nu$ representation in which states are wave functions $\tilde{\Psi}(\nu)$ and the operator measuring the physical volume of $\mathcal{V}$ is given by
\be\label{eq:V operator}
\hat{V}\tilde{\Psi}(\nu) \equiv 2\pi\Pl^2\gamma\abs{\hat{\nu}}\tilde\Psi(\nu)
=2\pi\Pl^2\gamma\abs{\nu}\tilde\Psi(\nu).
\ee
As before, the variable $b$ is canonically conjugate to $\nu$. However, unlike $\hat{\nu}$, the operator $\hat{b}$ is no longer well-defined in LQC and should be replaced by the holonomies, which can be promoted to operators with the action:
\be
\widehat{e^{\pm i\mubar c}}\tilde{\Psi}(\nu)\equiv
\widehat{e^{\pm i\lambda b}}\tilde{\Psi}(\nu)=\tilde{\Psi}(\nu\pm2\lambda).
\ee

The classical Hamiltonian constraint \eqnref{eqn:Cprime n} will be taken as our departing point for the quantum theory.
For the matter part, we adopt the ordinary Schr\"{o}dinger representation in which $\hat{\phi}$ is a multiplicative operator and $\hat{p}_\phi$ acts as a differential operator: i.e. $\hat{\phi}\tilde{\Psi}(\nu,\phi)=\phi\tilde{\Psi}(\nu,\phi)$ and $\hat{p}_\phi\tilde{\Psi}(\nu,\phi)=-i\hbar\partial_\phi\tilde{\Psi}(\nu,\phi)$.
Now we are ready to promote \eqnref{eqn:Cprime n} to the quantum Hamiltonian equation:
\be\label{eqn:LQC Hamiltonian eq nth}
\partial_\phi^2\tilde{\Psi}(\nu,\phi)
=-3\pi G \abs{\hat{\nu}}\,{\hat{b}_h^{(n)}}\abs{\hat{\nu}}\,{\hat{b}_h^{(n)}}
\tilde{\Psi}(\nu,\phi),
\ee
where ${\hat{b}_h^{(n)}}$ is defined as
\be\label{qen:bh n}
{\hat{b}_h^{(n)}}:=\frac{1}{\lambda}\sum_{k=0}^{n} \frac{(2k)!}{2^{2k}(k!)^2(2k+1)}\,
\widehat{\sin\lambda b}^{\ 2k+1}
\ee
and a particular factor ordering has been chosen to concur with the ordering adopted in \cite{Ashtekar:2007em}.
Additionally, physical states must lie in one of the irreducible representations (symmetric or antisymmetric) of the orientation reversal: $\Pi: \nu\longmapsto\Pi(\nu)=-\nu$, which is regarded as a large gauge transformation. Since there are no fermions in this model, $\tilde{\Psi}$ is assumed to be symmetric: $\Pi\tilde{\Psi}(\nu,\phi):=
\tilde{\Psi}(-\nu,\phi)=\tilde{\Psi}(\nu,\phi)$.

By taking this symmetry into account and expressing out the action of $\widehat{\sin\lambda b}$ explicitly, the Hamiltonian equation \eqnref{eqn:LQC Hamiltonian eq nth} is cast as
\ba\label{eqn:diffence equation}
&&\partial_\phi^2\tilde{\Psi}(\nu,\phi)=
-3\pi G \hat{\nu}\,{\hat{b}_h^{(n)}}\hat{\nu}\,{\hat{b}_h^{(n)}}
\tilde{\Psi}(\nu,\phi)\nn\\
&=&-\frac{3\pi G}{\lambda^2}\,
\nu\;
\sum_{k=0}^{n}\frac{(2k)!}{2^{2k}(k!)^2(2k+1)}\;
\sum_{k'=0}^{n}\frac{(2k')!}{2^{2k'}(k'!)^2(2k'+1)}\nn\\
&&
\;\times
\sum_{l_1,\,l_2=-(2n+1),-2n+1,\cdots}^{2n+1}
\left[
\frac{(-1)^{(2k+1-l_1)/2}(2k+1)!}
{(2i)^{2k+1}\left(\frac{2k+1-l_1}{2}\right)!\left(\frac{2k+1+l_1}{2}\right)!}
\right]
\left[
\frac{(-1)^{(2k'+1-l_2)/2}(2k'+1)!}
{(2i)^{2k'+1}\left(\frac{2k'+1-l_2}{2}\right)!\left(\frac{2k'+1+l_2}{2}\right)!}
\right]\nn\\
&&
\qquad\times
\left(\nu+l_1 \lambda+ l_2\lambda\right)
\tilde{\Psi}(\nu+2\,l_1\lambda+2\,l_2\lambda)\nn\\
&=:&-\hat{\Theta}^{(n)}_{(\nu)}\tilde{\Psi}(\nu,\phi),
\ea
which can be regarded as an ``evolution equation'' evolving the quantum state in the \emph{internal time} $\phi$. The kinematical Hilbert space is given by $\hil_\kin
=L^2(\mathbb{R}_\mathrm{Bohr},\abs{\nu}^{-1}dv_\mathrm{Bohr})\otimes
L^2(\mathbb{R},d\phi)$. The factor $1/\abs{\nu}$ is prescribed in the measure to have the operator $\hat{\Theta}^{(n)}_{(\nu)}$ self-adjoint.
For any order $n$, $\hat{\Theta}^{(n)}_{(\nu)}$ is a \emph{difference} operator in steps of $4\lambda$. For each $\epsilon\in[0,4\lambda)$, let $\hil_\kin^\epsilon$ denote the subspace of the kinematical Hilbert space $\hil_\kin$ with states whose support is restricted to the lattice points $\nu=\epsilon+4n\lambda$ (with $n\in \mathbb{Z}$). Each sector $\hil_\kin^\epsilon$ is preserved under the evolution equation; hence, there is \emph{superselection} among these sectors under dynamics. We can then focus on one of the sectors $\hil_\kin^\epsilon \oplus \hil_\kin^{4\lambda-\epsilon}$ if symmetry is also concerned. In this paper, without losing generality, we will focus on $\hil_\kin^{\epsilon=0}$ for simplicity (note $\hil_\kin^{\epsilon=0}
=\hil_\kin^{4\lambda}$). The treatment here should be straightforwardly extended for other sectors without much difficulty.

The physical Hilbert space $\hil_\phy$ can be obtained by applying the group averaging procedure to \eqnref{eqn:diffence equation}; the same strategies in \cite{Ashtekar:2007em} to construct the physical Hilbert space for the familiar $n=0$ case can be exactly followed. For generic $n$, again, the physical states satisfy the ``positive frequency'' square root of \eqnref{eqn:diffence equation}:
\be\label{eqn:positive frequency eq on nu}
-i\partial_\phi\tilde{\Psi}(\nu,\phi)=
\sqrt{\hat{\Theta}^{(n)}_{(\nu)}}\,\tilde{\Psi}(\nu,\phi)
\ee
and the \emph{physical scalar product} is given by
\be\label{eqn:physical scalar product on nu}
\pinner{\tilde{\Psi}_1}{\tilde{\Psi}_2}=
\frac{\lambda}{\pi}\sum_{\nu=4n\lambda}\frac{1}{\abs{\nu}}
\bar{\tilde{\Psi}}_1(\nu,\phi_0) \tilde{\Psi}_2(\nu,\phi_0),
\ee
where the right side is evaluated at any fixed instant $\phi_0$. The physical scalar product is independent of the choice of $\phi_0$ (i.e. conserved in internal time $\phi$) because $\hat{\Theta}^{(n)}_{(\nu)}$ (and hence its square root) is (positive definite and) self-adjoint with respect to $\pinner{\cdot}{\cdot}$ given above.
As in the case of $n=0$, the state with support at $\nu=0$ yields infinite norm and hence does not belong to the physical Hilbert space. This is a consequence of quantum dynamics since $\pinner{\cdot}{\cdot}$ is determined by the constraint operator.

Finally, we can define the complete set of Dirac observables. The first is the momentum $\hat{p}_\phi$ defined as
\be\label{eqn:Dirac op pphi on nu}
\hat{p}_\phi\tilde{\Psi}(\nu,\phi)
=-i\hbar\frac{\partial \tilde{\Psi}(\nu,\phi)}{\partial\phi}
\equiv\hbar\sqrt{\hat{\Theta}^{(n)}_{(\nu)}}\,\tilde{\Psi}(\nu,\phi)
\ee
and the second is a 1-parameter family of the ``relational'' observables $\hat{V}\vert_{\phi_0}$ defined as
\be\label{eqn:Dirac op V on nu}
\hat{V}\vert_{\phi_0}\tilde{\Psi}(\nu,\phi)
=2\pi\Pl^2\gamma\, \exp\left(i\sqrt{\hat{\Theta}^{(n)}_{(\nu)}}\,(\phi-\phi_0)\right)
\abs{\nu}\tilde{\Psi}(\nu,\phi_0),
\ee
which represents the volume at time $\phi_0$ by first freezing the solution $\tilde{\Psi}(\nu,\phi)$ at $\phi=\phi_0$, acting on it by the volume operator $\hat{V}$ defined in \eqnref{eq:V operator}, and evolving it from $\phi_0$ to $\phi$ by \eqnref{eqn:positive frequency eq on nu}. Obviously, both $\hat{p}_\phi$ and $\hat{V}\vert_{\phi_0}$ preserve the solutions to \eqnref{eqn:positive frequency eq on nu} and are self-adjoint with respect to \eqnref{eqn:physical scalar product on nu} since the square root of $\hat{\Theta}^{(n)}_{(\nu)}$ is self-adjoint.

To summarize, the physical Hilbert space $\hil_\phy$ (associated with the superselected sector $\hil_\kin^{\epsilon=0}$) consists of solutions $\tilde{\Psi}(\nu,\phi)$ to \eqnref{eqn:positive frequency eq on nu} (with support in the set $\{v=4n\lambda\}$) which are symmetric in $\nu$ and have finite norm with respect to \eqnref{eqn:physical scalar product on nu}. A complete set of Dirac observables are given by $\hat{p}_\phi$ and $\hat{V}\vert_{\phi_0}$ as defined in \eqnref{eqn:Dirac op pphi on nu} and \eqnref{eqn:Dirac op V on nu}, respectively.

\subsection{The $b$ and $x$ representations}\label{sec:b and x representations}
Since $\tilde{\Psi}(\nu,\phi)\in\hil_\kin^{\epsilon=0}$ has support on the ``\,lattice'' points $\{\nu=4n\lambda;n\in\mathbb{Z}\}$, the state $\ket{\Psi}$ in the $b$ representation gives the wave function
\be
\Psi(b,\phi):=\inner{b}{\Psi}=\sum_{\nu=4n\lambda}\inner{b}{\nu}\inner{\nu}{\Psi}
=\sum_{\nu=4n\lambda}e^{\frac{i}{2}\nu b} \,\tilde{\Psi}(\nu,\phi)
\ee
as the Fourier transform of $\tilde{\Psi}(\nu,\phi)$. It follows that $\Psi(b+\pi/\lambda,\phi)=\Psi(b,\phi)$ is a periodic function and the inverse Fourier transform reads as
\be
\tilde{\Psi}(\nu,\phi):=\inner{\nu}{\Psi}
=\frac{\lambda}{\pi}\int_0^{\pi/\lambda}db\,\inner{\nu}{b}\inner{b}{\Psi}
=\frac{\lambda}{\pi}\int_0^{\pi/\lambda}db\,
e^{-\frac{i}{2}\nu b} \,{\Psi}(b,\phi).
\ee
Additionally, the symmetry requirement $\tilde{\Psi}(\nu,\phi)=\tilde{\Psi}(-\nu,\phi)$ is equivalent to $\Psi(b,\phi)=\Psi(-b,\phi)$. In $b$ representation, $\hat{\nu}$ and $\widehat{\sin\lambda b}$ act as
\be
\hat{\nu}\,\Psi(b,\phi)=-2\,i\frac{\partial}{\partial b}\Psi(b,\phi)
\ee
and
\be
\widehat{\sin\lambda b}\,\Psi(b,\phi)=(\sin\lambda b) \Psi(b,\phi).
\ee

Following the treatment used in \cite{Ashtekar:2007em}, set $\tilde{\chi}(\nu,\phi):=\lambda/(\pi\nu)\tilde{\Psi}(\nu,\phi)$ and let $\chi(b,\phi):=\sum_{\nu=4n\lambda}e^{i\nu b/2}\tilde{\chi}(\nu,\phi)$ be the Fourier transform of $\tilde{\chi}(\nu,\phi)$. Then, on $\chi(b,\phi)$, the Hamiltonian constraint \eqnref{eqn:diffence equation} becomes
\be\label{eqn:master eq on b}
\partial_\phi^2\,\chi(b,\phi)=12\pi G
\left(b_h^{(n)}\partial_b\right)^2\chi(b,\phi)
=:-\hat{\Theta}_{(b)}^{(n)}\chi(b,\phi).
\ee
Also note that the symmetry requirement $\tilde{\Psi}(-\nu,\phi)=\tilde{\Psi}(\nu,\phi)$ is now translated to $\tilde{\chi}(-\nu,\phi)=-\tilde{\chi}(\nu,\phi)$ and $\chi(-b,\phi)=-\chi(b,\phi)$.

In the $b$ representation, the physical scalar product \eqnref{eqn:physical scalar product on nu} takes the form
\be\label{eqn:physical scalar product on b}
\pinner{\chi_1}{\chi_2}
=\int_0^{\pi/\lambda}db\,
\bar{\chi}_1(b,\phi_0)\abs{\hat{\nu}}\chi_2(b,\phi_0),
\ee
where $\abs{\hat{\nu}}$ is the positive square root of the self-adjoint operator $\hat{\nu}^2=(-2i\partial_b)^2$ on $\hil_\kin^{\epsilon=0}=L^2([0,\pi/\lambda),db)$.\footnote{To arrive at \eqnref{eqn:physical scalar product on b} from \eqnref{eqn:physical scalar product on nu}, we have to apply the formula for the Fourier transform of a \emph{Dirac comb}:
\be
\frac{1}{T}\sum_{k=-\infty}^{\infty}\delta\left(f-\frac{k}{T}\right)
=\sum_{n=-\infty}^{\infty}e^{-i\,2n\pi f T},\nn
\ee
which leads to
\ba
\sum_{\nu=4n\lambda}e^{-\frac{i}{2}\nu(b-b')}
&=& \frac{\pi}{\lambda}\sum_{k=-\infty}^{\infty}
\delta\left(b-b'-k\frac{\pi}{\lambda}\right)\nn\\
&=& \frac{\pi}{\lambda}\,\delta(b-b'),
\quad \text{if } b,b'\in[0,\pi/\lambda).\nn
\ea
This equation can also be used to show
\be
\Psi(b,\phi)=\frac{\pi}{\lambda}\,(-2i\partial_b)\chi(b,\phi)
\equiv \frac{\pi}{\lambda}\,\hat{\nu}\,\chi(b,\phi).\nn
\ee
}
The Dirac observables defined in \eqnref{eqn:Dirac op pphi on nu} and \eqnref{eqn:Dirac op V on nu} now read as
\be\label{eqn:Dirac op pphi on b}
\hat{p}_\phi\chi(b,\phi)
=-i\hbar\frac{\partial \chi(b,\phi)}{\partial\phi}
\equiv\hbar\sqrt{\hat{\Theta}_{(b)}^{(n)}}\,\chi(b,\phi)
\ee
and
\be\label{eqn:Dirac op V on b}
\hat{V}\vert_{\phi_0}\chi(b,\phi)
=2\pi\Pl^2\gamma\
\exp\left(i\sqrt{\hat{\Theta}_{(b)}^{(n)}}\,(\phi-\phi_0)\right)
\abs{\hat{v}}\,\chi(b,\phi_0).
\ee

To simplify the Hamiltonian constraint further, let us define the new variable $x$
\be
x=\int_\frac{\pi}{2\lambda}^{\,b} \frac{db'}{\sqrt{12\pi G}\,b_h^{(n)}\!(b')}
\qquad \text{for } b\in[0,{\pi}/{\lambda}),
\ee
such that $\sqrt{12\pi G}\,b_h^{(n)}\partial_b=\partial_x$ and $x\in(-\infty,\infty)$. For generic $b$, the corresponding new variable $x$ is then defined as $x(b)=x(b_0)$ with $b_0=b-n\,{\pi}/{\lambda}\in[0,{\pi}/{\lambda})$ for some $n\in \mathbb{Z}$.\footnote{By this periodic definition, the $\pi/\lambda$-periodic function $\chi(b,\phi)$ is well posed as $\chi(x,\phi):=\chi(b,\phi)$ with the new variable $x\in(-\infty,\infty)$. For generic $b$, we have $\sqrt{12\pi G}\,|b_h^{(n)}|\partial_b=\partial_x$ and the symmetry requirement $\chi(-b,\phi)=-\chi(b,\phi)$ now reads as $\chi(-x,\phi)=-\chi(x,\phi)$ in terms of the new variable $x$.}
Particularly, for the case of $n=0$, we have $b_h^{(n=0)}=\sin(\lambda b)/\lambda$ and
\be
x=\frac{1}{\sqrt{12\pi G}}\,\ln(\tan\frac{\lambda b}{2}).
\ee
For the case of $n=\infty$, it follows from \eqnref{qen:bh n} that $b_h:=b_h^{(\infty)}$ is a triangle wave with period $2\pi/\lambda$ and yields
\be\label{eqn:bh}
b_h:=b_h^{(\infty)}=
\left\{
\begin{array}{ccc}
b, & \text{  for} &  0\leq b \leq \frac{\pi}{2\lambda},\\
\frac{\pi}{\lambda}-b, & \text{  for} & \frac{\pi}{2\lambda}\leq b\leq \frac{\pi}{\lambda},
\end{array}
\right.
\ee
when restricted to $[0,\pi/\lambda]$;\footnote{Note that $b_h^{(n)}$ ($b_h$ included) is $2\pi/\lambda$-periodic and antisymmetric under $b\rightarrow-b$, but nevertheless the operator $(b_h^{(n)}\partial_b)^2$ and its positive square root are $\pi/\lambda$-periodic and symmetric under $b\rightarrow-b$. Thus, when acting on $\chi(b,\phi)$, the operator $(b_h^{(n)}\partial_b)^2$ and its positive square root do not alter the $\pi/\lambda$-period and antisymmetry of $\chi(b,\phi)$. In particular, $b_h$ is a triangle wave with period $2\pi/\lambda$ and agrees with $b$ only for $b\in[-\pi/(2\lambda),\pi/(2\lambda)]$. The distinction between $b_h$ and $b$ is an essential consequence of the intrinsic discreteness of LQC. That is, for any $\Psi\in\hil_\kin$, $\Psi(b,\phi)$ is an \emph{almost periodic} function of $b$, which becomes \emph{periodic} if restricted to $\hil_\kin^\epsilon$.} correspondingly, the new variable $x$ reads as
\be\label{eqn:x}
x=
\left\{
\begin{array}{lcc}
\frac{1}{\sqrt{12\pi G}}\ln\frac{2\lambda}{\pi}b, & \text{  for} &  0\leq b \leq \frac{\pi}{2\lambda},\\
-\frac{1}{\sqrt{12\pi G}}\ln\frac{2\lambda}{\pi}\left(\frac{\pi}{\lambda}-b\right),
& \text{  for} & \frac{\pi}{2\lambda}\leq b\leq \frac{\pi}{\lambda}.
\end{array}
\right.
\ee
For other cases of $n$, the expression of $x$ as a function of $b$ is much more complicated.

In terms of the new variable $x$, \eqnref{eqn:master eq on b} becomes the familiar Klein-Gordon equation:
\be\label{eqn:master eq on x}
\partial_\phi^2\chi(x,\phi)=\partial_x^2\chi(x,\phi)
=:-\hat{\Theta}_{(x)}^{(n)}\,\chi(x,\phi).
\ee
Again, the physical states can be taken to be positive frequency solutions to \eqnref{eqn:master eq on x}, i.e., solutions to
\be\label{eqn:positive frequency eq on x}
-i\partial_\phi\chi(x,\phi)=\sqrt{\hat{\Theta}_{(x)}^{(n)}}\,\chi(x,\phi).
\ee

If the initial data at the instant $\phi=\phi_0$ are given by
\be\label{eqn:Fourier series of chi}
\chi(x,\phi_0)=\frac{1}{\sqrt{2\pi}}\int_{-\infty}^\infty dk\, e^{-ikx}\tilde{\chi}(k),
\ee
then the physical state as the solution of \eqnref{eqn:positive frequency eq on x} takes the form
\ba\label{eqn:physical sol}
\chi(x,\phi)&=&\frac{1}{\sqrt{2\pi}} \int_{-\infty}^\infty dk\,
e^{-ikx+i\abs{k}(\phi-\phi_0)}\tilde{\chi}(k)\nn\\
&=&\frac{1}{\sqrt{2\pi}} \int_{-\infty}^0 dk\,
e^{-ik(\phi+x)}e^{ik\phi_0}\tilde{\chi}(k)
+\frac{1}{\sqrt{2\pi}} \int_0^{\infty} dk\,
e^{ik(\phi-x)}e^{-ik\phi_0}\tilde{\chi}(k)\nn\\
&=&\chi_L(\phi+x)+\chi_R(\phi-x).
\ea
As usual, the solutions can be decomposed into the left and right moving modes.
The group averaging procedure implies that the physical product is the standard scalar product for the Klein-Gordon equation:
\ba\label{eqn:physical scalar product on x}
\pinner{\chi_1}{\chi_2}&=&-i\int_{-\infty}^\infty dx
\left[\,
\bar{\chi}_1(x,\phi)\,\partial_\phi\chi_2(x,\phi)
-\left(\partial_\phi\bar{\chi}_1(x,\phi)\right)\chi_2(x,\phi)
\right],\nn\\
&=&2\int_{-\infty}^\infty dk\,\abs{k}\bar{\tilde{\chi}}_1(k)\tilde{\chi}_2(k),
\ea
which is conserved in the internal time $\phi$. Using $\partial_\phi\chi_L(\phi+x)=\partial_x\chi_L(\phi+x)$  and $\partial_\phi\chi_R(\phi-x)=-\partial_x\chi_R(\phi-x)$, it can be shown that the left and right moving sectors are mutually orthogonal, i.e., $\pinner{\chi_{1L}}{\chi_{2R}}=0$. This follows
\be\label{eqn:physical scalar product on x 2}
\pinner{\chi_1}{\chi_2}=-2i\int_{-\infty}^\infty dx
\left[\,
\bar{\chi}_{1L}(\phi+x)\,\partial_x\chi_{2L}(\phi+x)
-\bar{\chi}_{1R}(\phi-x)\,\partial_x\chi_{2R}(\phi-x)
\right].
\ee
Noting that the operator $\hat{\nu}=-2i\partial_b=-2i(12\pi G)^{-1/2}{b_h^{(n)}}^{-1}\partial_x$ is positive definite on the left moving sector and negative definite on the right moving sector, we then have
\ba
\pinner{\chi_1}{\chi_2}&=&\int_{-\infty}^\infty dx\,
\bar{\chi}_1(x,\phi)\, \widehat{\abs{-2i\partial_x}}\, \chi_2(x,\phi)\nn\\
&=&\int_0^{\pi/\lambda} db\,
\bar{\chi}_1(x(b),\phi)\, \widehat{\abs{-2i\partial_b}}\, \chi_2(x(b),\phi),
\ea
where $\widehat{\abs{-2i\partial_x}}$ and $\widehat{\abs{-2i\partial_b}}$ denote the positive parts of the self-adjoint operators $-2i\partial_x$ and $-2i\partial_b$ on $L^2(\mathbb{R},dy)$ and $L^2([0,\pi/\lambda),db)$, respectively. This equation agrees with \eqnref{eqn:physical scalar product on b}.

By \eqnref{eqn:Dirac op pphi on b} and \eqnref{eqn:Dirac op V on b}, in the $x$ representation, the Dirac observables act as
\be\label{eqn:Dirac op pphi on x}
\hat{p}_\phi\chi(x,\phi)
=-i\hbar\,\partial_\phi\chi(x,\phi)
=-i\hbar\,\partial_x\chi_L(\phi+x)+i\hbar\,\partial_x\chi_R(\phi-x)
\ee
and
\ba\label{eqn:Dirac op V on x}
\hat{V}\vert_{\phi_0}\chi(x,\phi)
&=&2\pi\Pl^2\gamma\
e^{i\sqrt{\hat{\Theta}_{(x)}^{(n)}}\,(\phi-\phi_0)}\,
\abs{\hat{v}}\,\chi(x,\phi_0)\\
&=&2\pi\Pl^2\gamma\
e^{i\sqrt{\hat{\Theta}_{(x)}^{(n)}}\,(\phi-\phi_0)}
\left[\,\hat{v}\chi_L(\phi_0+x)-\hat{v}\chi_R(\phi_0-x)\right]\nn\\
&=&-2i\frac{2\pi\Pl^2\gamma}{\sqrt{12\pi G}}\ e^{i\sqrt{\hat{\Theta}_{(x)}^{(n)}}\,(\phi-\phi_0)}
\left[{b_h^{(n)}\!(x)}^{-1}\partial_x\chi_L(\phi_0+x)
-{b_h^{(n)}\!(x)}^{-1}\partial_x\chi_R(\phi_0-x)\right]\nn\\
&=&-2i\frac{2\pi\Pl^2\gamma}{\sqrt{12\pi G}}
\left[{b_h^{(n)}\!(x\!+\!\phi\!-\!\phi_0)}^{-1}\partial_x\chi_L(\phi+x)
-{b_h^{(n)}\!(x\!-\!\phi\!+\!\phi_0)}^{-1}\partial_x\chi_R(\phi-x)\right].\nn
\ea

Taking into account the symmetry requirement $\chi(-x,\phi)=-\chi(x,\phi)$, we have
\be\label{eqn:symmetry of chi}
\chi(x,\phi)=\chi_L(\phi+x)+\chi_R(\phi-x)
=\frac{1}{\sqrt{2}}\left(F(\phi+x)-F(\phi-x)\right),
\ee
i.e. $\chi_L(x)=-\chi_R(x)=F(x)/\sqrt{2}$\,. Equivalently, this is to say $\tilde{\chi}(-k)=-\tilde{\chi}(k)$ in \eqnref{eqn:physical sol}. The full information in any physical state $\chi(x,\phi)$ is contained in $F$, and we can conveniently describe $\hil_\phy$ in terms of the positive frequency, left moving solutions $F(\phi+x)$ (or, equivalently, the right moving solutions $F(\phi-x)$), which are free of any symmetry requirement.\footnote{Note that although $F(x)$ is free of any symmetry requirement, $F(x)$ is not an arbitrary function but the function such that $F(\phi+x)$ is the left moving solution (or, equivalently, $F(\phi-x)$ is the right moving solution). The general solution is given by \eqnref{eqn:F(x)}.}
In terms of $F$, the physical scalar product \eqnref{eqn:physical scalar product on x 2} is simply written as
\be\label{eqn:physical scalar product on x 3}
\pinner{\chi_1}{\chi_2}=-2\,i\int_{-\infty}^\infty dx\,
\bar{F}_1(x)\,\partial_x F_2(x).
\ee
Equations \eqnref{eqn:Dirac op pphi on x} and \eqnref{eqn:Dirac op V on x} then give the matrix elements of the Dirac operators as
\ba\label{eqn:matrix element of pphi}
\pinner{\chi_1}{\hat{p}_\phi\,\chi_2}&=&
-2\hbar\int_{-\infty}^\infty dx
\left[\,
\bar{\chi}_{1L}(\phi+x)\,\partial_x^2\chi_{2L}(\phi+x)
+\bar{\chi}_{1R}(\phi-x)\,\partial_x^2\chi_{2R}(\phi-x)
\right]\nn\\
&=&
2\hbar\int_{-\infty}^\infty dx
\left[\,
\partial_x\bar{\chi}_{1L}(\phi+x)\,\partial_x\chi_{2L}(\phi+x)
+\partial_x\bar{\chi}_{1R}(\phi-x)\,\partial_x\chi_{2R}(\phi-x)
\right]\nn\\
&=&2\hbar\int_{-\infty}^\infty dx\,
\partial_x\bar{F}_1(x)\,\partial_x F_2(x)
\ea
and
\ba\label{eqn:matrix element of V}
\pinner{\chi_1}{\hat{V}\vert_{\phi_0}\chi_2}&=&
\frac{4(2\pi\Pl^2\gamma)}{\sqrt{12\pi G}}
\int_{-\infty}^\infty dx
\left[\,
\partial_x\bar{\chi}_{1L}(\phi_0+x)\,{b_h^{(n)}\!(x)}^{-1}\partial_x\chi_{2L}(\phi_0+x)
\right.\\
&&\qquad\qquad\qquad\qquad
\left.
+\,\partial_x\bar{\chi}_{1R}(\phi_0-x)\,{b_h^{(n)}\!(x)}^{-1}\partial_x\chi_{2R}(\phi_0-x)
\right]\nn\\
&=&\frac{2(2\pi\Pl^2\gamma)}{\sqrt{12\pi G}}
\int_{-\infty}^\infty dx\,
\partial_x\bar{F}_1(x)
\left[{b_h^{(n)}\!(x-\phi_0)}^{-1}\!\!+{b_h^{(n)}\!(\phi_0-x)}^{-1}\right]
\partial_x F_2(x).\nn
\ea

Similarly, for later use, we can also derive
\be\label{eqn:matrix element of pphi square}
\pinner{\chi_1}{\hat{p}_\phi^2\,\chi_2}
=\pinner{\hat{p}_\phi\,\chi_1}{\hat{p}_\phi\,\chi_2}
=-2i\hbar^2\int_{-\infty}^\infty dx\,
\partial_x\bar{F}_1(x)\,\partial_x^2 F_2(x)
\ee
and
\ba\label{eqn:matrix element of V square}
&&\pinner{\chi_1}{(\hat{V}\vert_{\phi_0})^2\chi_2}=
\pinner{\hat{V}\vert_{\phi_0}\chi_1}{\hat{V}\vert_{\phi_0}\chi_2}\nn\\
&=&\frac{-8i\pi G\hbar^2\gamma^2}{3}
\int_{-\infty}^\infty dx
\left(b_h^{(n)}\!(x-\phi_0)^{-1}\partial_x\bar{F}(x)\right)
\partial_x
\left(b_h^{(n)}\!(x-\phi_0)^{-1}\partial_x F(x)\right).
\ea

\subsection{Singularity resolution and the absolute upper bound on matter density}
\label{sec:singularity resolution}
We have noted for \eqnref{eqn:physical scalar product on nu} that the state with support at $\nu=0$ does not belong to the physical Hilbert space $\hil_\phy$. This by itself, however, does not mean that the singularity is avoided. (See \cite{Ashtekar:2007em} for more comments on weaker notions of singularity resolution.)

In fact, it is nontrivial to obtain necessary and sufficient conditions to characterize the occurrence of a singularity even in classical general relativity, and thus a generally applicable and satisfactory notion of ``singularity resolution'' is not available in quantum gravity. In simple situations, however, the notion of a singularity is unambiguous in the classical theory and a satisfactory notion of singularity resolution is possible in the quantum theory. A sensible notion is proposed in \cite{Ashtekar:2007em}: The singularity is said to be resolved in the quantum theory if the physical Hilbert space and a complete family of Dirac observables are available and the expectation values of these observables remain finite in the regime in which they become classically singular. In the model studied in this paper, the matter density diverges at the classical singularity. The singularity is then considered as resolved in the quantum theory if we can show that the expectation values of a complete set of Dirac observables including the matter density remain finite.\footnote{Note that the exclusion of the state with support at $v=0$ does not imply that the expectation values of the matter density would be bounded from above on the physical Hilbert space $\hil_\phy$, because $\hat{p}_\phi$ does not have an upper bound on $\hil_\phy$.}

For the case of $n=0$, it has been shown in \cite{Ashtekar:2007em} that the matter density is bounded by an upper bound as far as two natural notions representing the mean value of the matter density are considered. We will show that the same is also true for the cases of generic $n$, while the upper bound is modified by a numerical factor involving $n$.

First, let us define the mean value of the matter density (as a function of the internal time $\phi$) as
\be\label{eqn:density notion 1}
\rho_\phi(\phi):=
\frac{\pinner{\chi}{\hat{p}_\phi\,\chi}^{\,2}}
{2\,\pinner{\chi}{\hat{V}\vert_{\phi}\chi}^{\,2}}
\ee
and use it to represent the measurement of matter density.
From \eqnref{eqn:matrix element of pphi} and \eqnref{eqn:matrix element of V}, it follows
\be
\rho_\phi(\phi)=
\frac
{
\frac{3}{2\pi G\gamma^2}
\left[\int_{-\infty}^\infty dx\,
\abs{\partial_x F(x)}^2\right]^2
}
{
\left[\int_{-\infty}^\infty dx\,
\abs{\partial_x F(x)}^2
\left({b_h^{(n)}\!(x-\phi)}^{-1}\!\!+{b_h^{(n)}\!(\phi-x)}^{-1}\right)
\right]^2
}\,.
\ee
Noting that \eqnref{qen:bh n} gives
\be\label{eqn:bh n upper bound}
b_h^{(n)}\leq
\frac{1}{\lambda}\sum_{k=0}^{n} \frac{(2k)!}{2^{2k}(k!)^2(2k+1)}
=:\frac{\mathfrak{F}_n}{\lambda},
\ee
we then have
\be\label{eqn:rho bounded 1}
\rho_\phi(\phi)\leq
\frac{3}{2\pi G\gamma^2}
\frac{\int_{-\infty}^\infty dx \abs{\partial_x F}^2}
{(2\lambda\, \mathfrak{F}_n^{-1})^2\int_{-\infty}^\infty dx \abs{\partial_x F}^2}
=3\,\mathfrak{F}_n^2\rho_\mathrm{Pl}
=:\rho_\mathrm{sup}^{(n)},
\ee
where the Planckian density is defined in \eqnref{eqn:Planckian density}.

This shows that the mean density $\rho_\phi(\phi)$ is bounded from above by an absolute upper bound $\rho_\mathrm{sup}^{(n)}$, which exactly coincides with the critical density $\rho_\mathrm{crit}^{(n)}$ obtained at the level of heuristic effective dynamics \cite{Chiou:2009hk}. Also note that
\be\label{eqn:rho sup}
3\rho_\mathrm{Pl}=
\rho_\mathrm{sup}^{(0)} < \rho_\mathrm{sup}^{(1)} < \cdots < \rho_\mathrm{sup}^{(\infty)}=\frac{3\pi^2}{4}\rho_\mathrm{Pl}.
\ee
That is, the upper bound $\rho_\mathrm{sup}^{(n)}$ increases with $n$, but remains finite in the Planckian regime even if all orders of holonomy corrections are included.

Second, we can use the expectation value of the density operator in place of $\rho_\phi$ used above to represent the mean density. To have a self-adjoint operator, let us define the density operator at the instant $\phi$ as
\be
\hat{\rho}_\phi(\phi):=\frac{1}{2}
(\hat{V}\vert_\phi)^{-1}\,\hat{p}_\phi^2\,(\hat{V}\vert_\phi)^{-1}.
\ee
Analogous to \eqnref{eqn:Dirac op V on x}, the inverse volume operator $\hat{V}\vert_{\phi_0}$ at the instant $\phi_0$ is defined as
\ba
(\hat{V}\vert_{\phi_0})^{-1}\chi(x,\phi)
&:=&(2\pi\Pl^2\gamma)^{-1}
e^{i\sqrt{\hat{\Theta}_{(x)}^{(n)}}\,(\phi-\phi_0)}\,
|\widehat{v^{-1}}|\,\chi(x,\phi_0)\\
&=&(2\pi\Pl^2\gamma)^{-1}
e^{i\sqrt{\hat{\Theta}_{(x)}^{(n)}}\,(\phi-\phi_0)}
\left[\widehat{v^{-1}}\,\chi_L(\phi_0+x)-\widehat{v^{-1}}\,\chi_R(\phi_0-x)\right]\nn\\
&=&\frac{i\sqrt{12\pi G}}{4\pi\Pl^2\gamma}\ e^{i\sqrt{\hat{\Theta}_{(x)}^{(n)}}\,(\phi-\phi_0)}
\left[\partial_x^{-1}\left({b_h^{(n)}}\chi_L(\phi_0+x)\right)
-\partial_x^{-1}\left({b_h^{(n)}}\chi_R(\phi_0-x)\right)\right],\nn
\ea
where $|\widehat{v^{-1}}|$ is the inverse of $\abs{\hat{v}}$,\footnote{Note that $|\widehat{v^{-1}}|$ is not densely defined on $\hil_\kin$, but nevertheless it is well defined on $\hil_\phy$ since the state with support at $\nu=0$ is excluded on $\hil_\phy$.} and $\partial_x^{-1}$ is the inverse of $\partial_x$.\footnote{More precisely, when acting on $\chi(x,\phi)$, the operator $\partial_x^{-1}$ is given by
\be
\partial_x^{-1}\chi(x,\phi)
=\frac{1}{\sqrt{2\pi}}\int_{-\infty}^\infty dk\, \frac{1}{-ik}\,
e^{-ikx+i\abs{k}(\phi-\phi_0)}\,\tilde{\chi}(k).\nn
\ee} Consequently, by \eqnref{eqn:Dirac op pphi on x}, it follows
\ba
&&\hat{p}_\phi\,
(\hat{V}\vert_{\phi_0})^{-1}\chi(x,\phi)
=\frac{\hbar\sqrt{12\pi G}}{4\pi\Pl^2\gamma}\ e^{i\sqrt{\hat{\Theta}_{(x)}^{(n)}}\,(\phi-\phi_0)}
\left[b_h^{(n)}\!(x)\chi_L(\phi_0+x)
+b_h^{(n)}\!(x)\chi_R(\phi_0-x)\right]\nn\\
&&\qquad\qquad=\frac{\hbar\sqrt{12\pi G}}{4\pi\Pl^2\gamma}
\left[b_h^{(n)}\!(x\!+\!\phi\!-\!\phi_0)\chi_L(\phi+x)
+b_h^{(n)}\!(x\!-\!\phi\!+\!\phi_0)\chi_R(\phi-x)\right].
\ea
By \eqnref{eqn:physical scalar product on x 3}, we then have
\ba
&&\pinner{\chi}{(\hat{V}\vert_{\phi_0})^{-1}\,\hat{p}_\phi^2\,
(\hat{V}\vert_{\phi_0})^{-1}\chi}
=\pinner{\hat{p}_\phi(\hat{V}\vert_{\phi_0})^{-1}\chi}
{\hat{p}_\phi(\hat{V}\vert_{\phi_0})^{-1}\chi}\nn\\
&=&\frac{-6i}{4\pi G\gamma^2}\int_{-\infty}^\infty dx\,
\left(b_h^{(n)}\!(x+\phi-\phi_0)\,\bar{F}(\phi+x)\right)
\partial_x\left(b_h^{(n)}\!(x+\phi-\phi_0)\, F(\phi+x)\right)\nn\\
&=&\frac{-6i}{4\pi G\gamma^2}\int_{-\infty}^\infty dx
\left(b_h^{(n)}\!(x-\phi_0)\, \bar{F}(x)\right)
\partial_x
\left(b_h^{(n)}\!(x-\phi_0)\, F(x)\right).
\ea
Therefore, the expectation value of the density operator yields
\ba\label{eqn:rho bounded 2}
\langle\hat{\rho}_\phi(\phi)\rangle
&:=&
\frac{\pinner{\chi}{(\hat{V}\vert_\phi)^{-1}\,\hat{p}_\phi^2\,
(\hat{V}\vert_\phi)^{-1}\chi}}
{2\pinner{\chi}{\chi}}\nn\\
&=&\frac{3}{8\pi G\gamma^2}
\frac{\pinner{\hat{b}_h^{(n)}\!(x-\phi)\chi}{\hat{b}_h^{(n)}\!(x-\phi)\chi}}
{\pinner{\chi}{\chi}}\nn\\
&\leq&\frac{3}{8\pi G\gamma^2}
\frac{(\lambda^{-1}\mathfrak{F}_n)^2\pinner{\chi}{\chi}}{\pinner{\chi}{\chi}}
=3\,\mathfrak{F}_n^2\rho_\mathrm{Pl}
=:\rho_\mathrm{sup}^{(n)},
\ea
where \eqnref{eqn:bh n upper bound} has been used again. This shows that, as the second notion of the mean density is regarded, the matter density is again bounded from above by the same upper bound $\rho_\mathrm{sup}^{(n)}$.

To summarize, both $\rho_\phi$ and $\langle\hat{\rho_\phi}\rangle$ can be used to represent the physical measurement of the matter density, and both of them are bounded from above by the same absolute upper bound $\rho_\mathrm{sup}^{(n)}$, which equals the critical density $\rho_\mathrm{crit}^{(n)}$ obtained at the level of heuristic effective dynamics. Therefore, we have shown that, at any order $n$, the singularity is resolved as far as the above two notions of the matter density measurement are concerned. Moreover, the singularity resolution holds for any arbitrary physical states, not only restricted to the states which are semiclassical at late times.

\subsection{Quantum bounce and the bouncing scenario}
\label{sec:bouncing scenario}
We have shown that the matter density is bounded from above in both senses of the matter density measurement. For the case of $n=0$, it can further be shown that, for any arbitrary (not necessarily semiclassical) states, the expectation value of the volume operator $\langle\hat{V}\vert_\phi\rangle$ follows the bouncing trajectory and the matter density at the bouncing epoch can come arbitrarily close to $\rho_{\sup}^{(n=0)}$ \cite{Ashtekar:2007em}. In the following, we will prove that the same conclusion can also be drawn for the case of $n=\infty$. (For the case of generic $n$, we expect the same but the algebra is much more complicated to give a rigorous proof.)

Equations \eqnref{eqn:bh} and \eqnref{eqn:x} yield
\be\label{eqn:bh of x}
b_h(x):=b_h^{(\infty)}(x)=
\left\{
\begin{array}{lcr}
\frac{\pi}{2\lambda}\,e^{\sqrt{12\pi G}\,x}, & \text{  for} &  -\infty\leq x \leq 0,\\
\frac{\pi}{2\lambda}\,e^{-\sqrt{12\pi G}\,x}, & \text{  for} & 0\leq x\leq \infty.
\end{array}
\right.
\ee
Taking this into \eqnref{eqn:matrix element of V} then gives
\ba\label{eqn:bouncing sol}
\langle\hat{V}\vert_\phi\rangle
&\equiv&\pinner{\chi}{\hat{V}\vert_\phi\chi}\nn\\
&=&
\frac{16\gamma\Pl^2\lambda}{\sqrt{12\pi G}}
\left(
\int_{-\infty}^\phi dx\, \abs{\partial_x F(x)}^2\, e^{-\sqrt{12\pi G}\,(x-\phi)}
+
\int_\phi^{\infty} dx\, \abs{\partial_x F(x)}^2\, e^{\sqrt{12\pi G}\,(x-\phi)}
\right)\nn\\
&=&
V_+(\phi)\,e^{\sqrt{12\pi G}\,\phi}
+V_-(\phi)\,e^{-\sqrt{12\pi G}\,\phi},
\ea
where
\begin{subequations}\label{eqn:V+-}
\ba
V_+(\phi)&=&\frac{16\gamma\Pl^2\lambda}{\sqrt{12\pi G}}
\int_{-\infty}^\phi dx\, \abs{\partial_x F(x)}^2\, e^{-\sqrt{12\pi G}\,x}
=:\frac{16\gamma\Pl^2\lambda}{\sqrt{12\pi G}}
\int_{-\infty}^\phi\!dx\,v_+(x),\\
V_-(\phi)&=&\frac{16\gamma\Pl^2\lambda}{\sqrt{12\pi G}}
\int_\phi^{\infty} dx\, \abs{\partial_x F(x)}^2\, e^{+\sqrt{12\pi G}\,x}
=:\frac{16\gamma\Pl^2\lambda}{\sqrt{12\pi G}}
\int_\phi^\infty\!dx\,v_-(x).
\ea
\end{subequations}
Similarly, taking \eqnref{eqn:bh of x} into \eqnref{eqn:matrix element of V square} gives
\be\label{eqn:bouncing sol of V square}
\langle(\hat{V}\vert_\phi)^2\rangle
\equiv\pinner{\chi}{(\hat{V}\vert_{\phi})^2\chi}
=
W_+(\phi)\,e^{2\sqrt{12\pi G}\,\phi}
+W_-(\phi)\,e^{-2\sqrt{12\pi G}\,\phi},
\ee
where
\begin{subequations}\label{eqn:W+-}
\ba
W_+(\phi)&=&\frac{32\,G\hbar^2\gamma^2\lambda^2}{3\pi}
\int_{-\infty}^\phi dx
\frac{1}{i}\left[e^{-\sqrt{12\pi G}\,x}\,\partial_x\bar{F}(x)\right]
\partial_x
\left[e^{-\sqrt{12\pi G}\,x}\,\partial_xF(x)\right]\nn\\
&=:&\frac{32\left(\Pl^2\gamma\lambda\right)^2}{3\pi G}
\int_{-\infty}^\phi dx\, w_+(x)\\
W_-(\phi)&=&\frac{32\,G\hbar^2\gamma^2\lambda^2}{3\pi}
\int_\phi^\infty dx
\frac{1}{i}\left[e^{+\sqrt{12\pi G}\,x}\,\partial_x\bar{F}(x)\right]
\partial_x
\left[e^{+\sqrt{12\pi G}\,x}\,\partial_xF(x)\right]\nn\\
&=:&\frac{32\left(\Pl^2\gamma\lambda\right)^2}{3\pi G}
\int_\phi^\infty dx\, w_-(x).
\ea
\end{subequations}

As long as $F(x)$ is smooth, the trajectories of $\langle\hat{V}\vert_\phi\rangle$, $\langle(\hat{V}\vert_\phi)^2\rangle$, and thus $(\Delta V\vert_\phi)^2 :=\langle(\hat{V}\vert_\phi)^2\rangle-\langle\hat{V}\vert_\phi\rangle$ are all smooth functions of $\phi$, despite the kink of the function $b_h(x)$ at $x=0$. This tells us that the abrupt kink of the solution obtained in \cite{Chiou:2009hk} at the level of heuristic effective dynamics is only an artifact, which is smeared by the quantum fluctuations at the level of quantum theory.
Furthermore, for the physical states which are highly semiclassical, the integrand in \eqnref{eqn:V+-} and (the real part of ) the integrand in \eqnref{eqn:W+-} are appreciable only for some localized intervals and diminish rapidly away from the intervals; i.e. $v_\pm(x)\simeq0$ if $x\not\in(x_v^-,x_v^+)$ and (the real part of) $w_\pm(x)\simeq0$ if $x\not\in(x_w^-,x_w^+)$ for some $x_v^\pm$ and $x_w^\pm$. Consequently, we have $V_+(\phi>x_v^+)\simeq V_+(\infty)=:V_+$ and $V_-(\phi>x_v^+)\simeq V_-(\infty)=0$; $V_-(\phi<x_v^-)\simeq V_-(-\infty)=:V_-$ and $V_+(\phi<x_v^-)\simeq0$; $W_+(\phi>x_w^+)\simeq W_+(\infty)=:W_+$ and $W_-(\phi>x_w^+)\simeq0$; as well as $W_-(\phi<x_w^-)\simeq W_-(-\infty)=:W_-$ and $W_+(\phi<x_w^-)\simeq0$. This leads to
\begin{subequations}\label{eqn:asymptotic sol}
\ba
\langle\hat{V}\vert_\phi\rangle
&\simeq&
\left\{
\begin{array}{lcr}
V_+\,e^{\sqrt{12\pi G}\,\phi}, & \text{  for} & \phi>x_v^+,\\
V_-\,e^{-\sqrt{12\pi G}\,\phi}, & \text{  for} & \phi<x_v^-,
\end{array}
\right.\\
\langle(\hat{V}\vert_\phi)^2\rangle
&\simeq&
\left\{
\begin{array}{lcr}
W_+\,e^{2\sqrt{12\pi G}\,\phi}, & \text{  for} & \phi>x_w^+,\\
W_-\,e^{-2\sqrt{12\pi G}\,\phi}, & \text{  for} & \phi<x_w^-,
\end{array}
\right.
\ea
\end{subequations}
with $V_\pm$ and $W_\pm$ being constants.\footnote{Note that while $v_\pm(x)$ is real, $w_\pm(x)$ is complex in general. However, the contributions of the imaginary part of $w_\pm(x)$ should exactly cancel out in \eqnref{eqn:bouncing sol of V square} since $\langle(\hat{V}\vert_\phi)^2\rangle$ is real.}
Therefore, in the distant future and past (i.e. $\phi>\max(x_v^+,x_w^+)$ or $\phi<\min(x_v^-,x_w^-)$, respectively), $\langle\hat{V}\vert_\phi\rangle$ follows the classical trajectory with constant relative uncertainty spread (i.e. $(\Delta V\vert_\phi)^2/\langle\hat{V}\vert_\phi\rangle^2\simeq (W_\pm-V_\pm^2)/V_\pm^2$). In other words, \eqnref{eqn:asymptotic sol} gives exactly the same behavior as the WDW theory in the far future and past. Thus, the physical solution \eqnref{eqn:bouncing sol} gives rise to the bouncing scenario in which two WDW solutions (expanding and contracting) are bridged together through a transition phase of the quantum bounce. Additionally, the more the integrands $v_\pm(x)$ and $w_\pm^\mathrm{Re}(x)$ are localized, the more abrupt the transition phase is. In principle, the transition phase can be arbitrarily brief if $v_\pm(x)$ and $w_\pm^\mathrm{Re}(x)$ are extremely sharp.

To understand the dynamics in more detail, let us study the evolution of Dirac observables for coherent (semiclassical) states explicitly. According to \eqnref{eqn:physical sol} and \eqnref{eqn:symmetry of chi}, the general solution of $F(x)$ is given by
\ba\label{eqn:F(x)}
F(x)&=&\sqrt{2}\,\chi_L(x)=-\sqrt{2}\,\chi_R(x)\nn\\
&=&-\frac{1}{\sqrt{2\pi}}\int_{-\infty}^0 dk\, e^{-ikx}\tilde{F}(k)
=\frac{1}{\sqrt{2\pi}}\int_0^\infty dk\, e^{+ikx}\tilde{F}(k),
\ea
where, without losing generality, the offset constant $\phi_0$ in \eqnref{eqn:physical sol} is set to $\phi_0=0$.\footnote{As we will see shortly, this choice will set the epoch of the quantum bounce at $\phi=0$.}
To construct a coherent state, we choose the Fourier amplitudes to be
\be
\tilde{F}(k)=-\sqrt{2}\,\tilde{\chi}(k)
=
\left\{
\begin{array}{lcr}
\frac{e^{\frac{-(k-k^*)^2}{(2\sigma)^2}}}{\sqrt{2}\,(2\pi\sigma^2)^{1/4}\sqrt{k_*}}\,, & \text{  for} &  k>0,\\
-\frac{e^\frac{-(k+k^*)^2}{(2\sigma)^2}}{\sqrt{2}\,(2\pi\sigma^2)^{1/4}\sqrt{k_*}}\,, & \text{  for} & k<0,
\end{array}
\right.
\ee
where $\tilde{F}(k)=-\tilde{F}(-k)$ and $\tilde{F}(k)$ takes the form of a Gaussian distribution centered at $k_*$ ($k_*>0$) with $\sigma$ representing the width.
In case of highly semiclassical states where the Gaussian distribution is very sharp with $\sigma\ll k_*$, we can accurately approximate \eqnref{eqn:F(x)} to
\be\label{eqn:coherent state}
F(x)
\approx
\frac{1}{\sqrt{2\pi}}\int_{-\infty}^\infty dk\, e^{+ikx}
\frac{e^{\frac{-(k-k^*)^2}{(2\sigma)^2}}}{\sqrt{2}\,(2\pi\sigma^2)^{1/4}\sqrt{k_*}}
=
\frac{\sqrt{\sigma}}{(2\pi)^{1/4}\sqrt{k_*}}\
e^{-\sigma^2x^2+ik_*x}.
\ee
By \eqnref{eqn:physical scalar product on x 3}, \eqnref{eqn:matrix element of pphi} and \eqnref{eqn:matrix element of pphi square}, it is straightforward to show that
$\pinner{\chi}{\chi}\approx1$ and
\begin{subequations}\label{eqn:expectation and uncertainty of pphi}
\ba
\label{eqn:expectation of pphi}
\langle\hat{p}_\phi\rangle
&\equiv&
\pinner{\chi}{\hat{p}_\phi\chi}
\approx\hbar\, k_* \left(1+(\sigma/k_*)^2\right),\\
\langle\hat{p}_\phi^2\rangle
&\equiv&
\pinner{\chi}{\hat{p}_\phi^2\,\chi}
\approx\hbar^2k_*^2\left(1+3(\sigma/k_*)^2\right),\\
(\Delta p_\phi)^2
&\equiv&\langle\hat{p}_\phi^2\rangle-\langle\hat{p}_\phi\rangle^2
\approx\hbar^2\sigma^2\left(1-(\sigma/k_*)^2\right).
\ea
\end{subequations}
That is, the coherent state given by \eqnref{eqn:coherent state} is normalized and the parameters $k_*$ and $\sigma$ are associated with the expectation value and the uncertainty spread of $\hat{p}_\phi$. With $\sigma\ll k_*$, \eqnref{eqn:expectation and uncertainty of pphi} then yields $\langle\hat{p}_\phi\rangle\approx \hbar k_*$ and $(\Delta p_\phi)^2\approx \hbar^2\sigma^2$.

Given with \eqnref{eqn:coherent state}, the integrand in \eqnref{eqn:V+-} reads as
\be\label{eqn:v+-}
v_\pm(x):=\abs{\partial_x F(x)}^2 e^{\mp\sqrt{12\pi G}\,x}
=\frac{\sigma}{\sqrt{2\pi}k_*}
\left(4\sigma^4x^2+k_*^2\right) e^{-2\sigma^2x^2\mp\sqrt{12\pi G}\,x}
\ee
and the integrand in \eqnref{eqn:W+-} reads as
\ba\label{eqn:w+-}
w_\pm(x)&:=&-i
\left[e^{\mp\sqrt{12\pi G}\,x}\,\partial_x\bar{F}(x)\right]
\partial_x
\left[e^{\mp\sqrt{12\pi G}\,x}\,\partial_xF(x)\right]\nn\\
&=&
\frac{\sigma}{\sqrt{2\pi}k_*}
e^{-2\sigma^2x^2\mp2\sqrt{12\pi G}\,x}
\bigg[
k_*\left(k_*^2+2\,\sigma^2+4\sigma^4x^2\right)\nn\\
&&\qquad\qquad
+\,i\left[
k_*^2\left(2\,\sigma^2x\pm\sqrt{12\pi G}\right)
+4\,\sigma^4x\left(-1\pm\sqrt{12\pi G}\,x+2\,\sigma^2x^2\right)
\right]
\bigg]\nn\\
&=:&w_\pm^\mathrm{Re}(x)+i\,w_\pm^\mathrm{Im}(x).
\ea
Note that for the imaginary part of $w_\pm(x)$ yields
\begin{subequations}\label{eqn:imaginary part of w}
\ba
\int_{-\infty}^\phi dx\, w_+^\mathrm{Im}(x)&=&
-\frac{\sigma}{2\sqrt{2\pi}k_*}\,e^{-2\sigma^2\phi^2-2\sqrt{12\pi G}\,\phi} \left(k_*^2+4\,\sigma^4\phi^2\right),\\
\int_\phi^\infty dx\, w_-^\mathrm{Im}(x)&=&
+\frac{\sigma}{2\sqrt{2\pi}k_*}\,e^{-2\sigma^2\phi^2+2\sqrt{12\pi G}\,\phi}
\left(k_*^2+4\,\sigma^4\phi^2\right).
\ea
\end{subequations}
The contributions of the imaginary part of $w_\pm(x)$ cancel out exactly in \eqnref{eqn:bouncing sol of V square} as expected. We can simply ignore $w_\pm^\mathrm{Im}(x)$ and replace $w_\pm(x)$ with the real part $w_\pm^\mathrm{Re}(x)$.

\begin{figure}
\begin{picture}(470,140)(0,0)

\put(18,15)
{
\scalebox{0.86}{\includegraphics{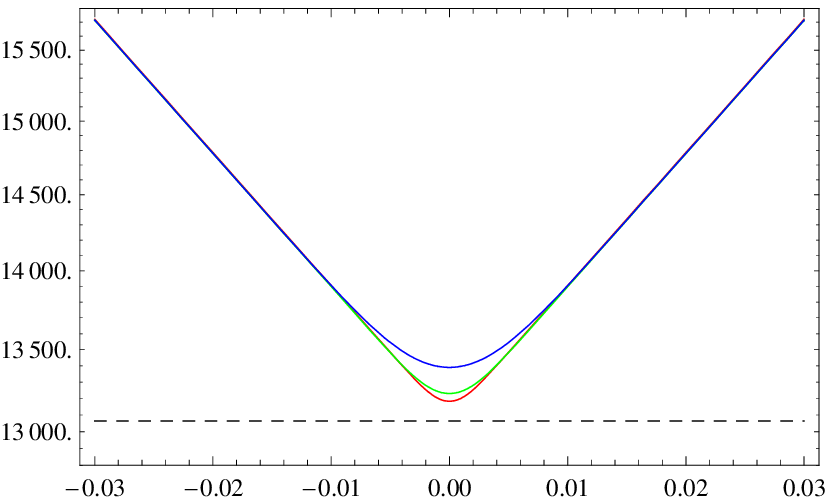}}
}
\put(0,130){\textbf{(a)}}
\put(108,5){\footnotesize $\phi$ ($G^{-1/2}$)}
\put(3,40){\rotatebox{90}{$\langle\hat{V}\vert_\phi\rangle$\; ($\gamma\Pl^2\lambda$)}}

\put(258,15)
{
\scalebox{0.85}{\includegraphics{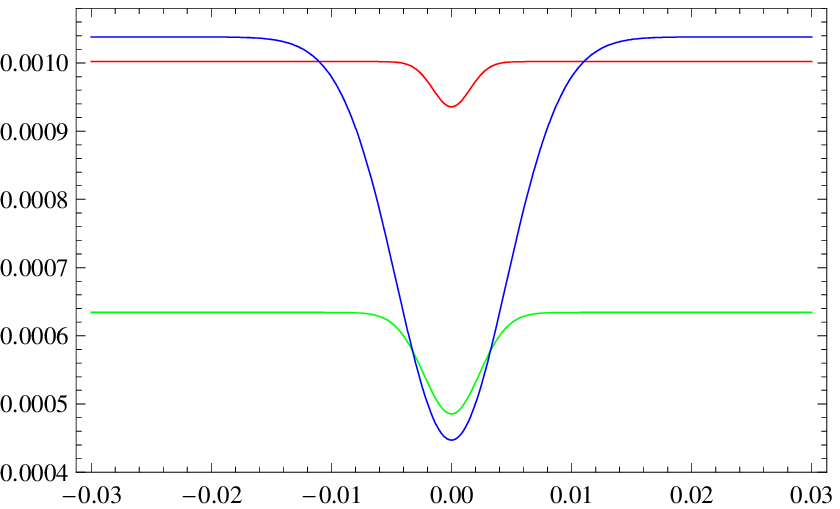}}
}
\put(238,130){\textbf{(b)}}
\put(348,5){\footnotesize $\phi$ ($G^{-1/2}$)}
\put(243,35){\rotatebox{90}{$(\Delta V\vert_\phi)^2/\langle\hat{V}\vert_\phi\rangle^2$}}
\put(450,130){\scriptsize i}
\put(449,60){\scriptsize ii}
\put(448,122){\scriptsize iii}

\end{picture}
\caption{The evolution of coherent states with $k_*=2.\times10^3\sqrt{8\pi G}$ and (i) $\sigma=0.01k_*$, (ii) $\sigma=0.02k_*$, and (iii) $\sigma=0.03k_*$. \textbf{(a)} The expectation value of volume $\langle\hat{V}\vert_\phi\rangle$ as a function of $\phi$ in the logarithmic scale. Two classical trajectories (straight lines in the logarithmic scale) are bridged through the quantum bounce.
The value of $V_\mathrm{bounce}$ decreases as $\sigma$ increases as suggested in \eqnref{eqn:Vbounce}. The value that corresponds to $\rho_{\sup}^{(\infty)}$ is indicated by the dashed line and sets the absolute lower bound for $V_\mathrm{bounce}$ as suggested in \eqnref{eqn:density at bounce}. \textbf{(b)} Relative uncertainty spread $(\Delta V\vert_\phi)^2/\langle\hat{V}\vert_\phi\rangle^2$, which asymptotes to the constant given by \eqnref{eqn:limiting relative spread part a}. Notice that as $\sigma$ is larger, the transition phase (during which the relative uncertainty spread deviates from the asymptotic constant) is briefer, in agreement with the discussion in the paragraph after \eqnref{eqn:imaginary part of w}.}\label{fig:evolution}
\end{figure}

Equations \eqnref{eqn:v+-} and \eqnref{eqn:w+-} show that both $v_\pm(x)$ and $w_\pm^\mathrm{Re}(x)$ are localized bumps with widths which decrease as $\sigma$ increases. This confirms what we have just discussed prior to \eqnref{eqn:asymptotic sol}. For a given $k_*$, the bigger $\sigma$ is, the narrower the bumps are and thus the briefer the transition phase of the quantum bounce is, as can been seen in \figref{fig:evolution}.

Substituting \eqnref{eqn:v+-} into \eqnref{eqn:V+-} and (the real part of) \eqnref{eqn:w+-} into \eqnref{eqn:W+-}, we have
\ba
\label{eqn:Vpm of phi}
V_\pm(\phi)&=&
\frac{4\gamma\Pl^2\lambda}{\sqrt{12\pi G}\,k_*}
\Bigg[
\frac{\sqrt{2}\,\sigma}{\sqrt{\pi}}\,e^{\mp2\phi(\sqrt{3\pi G}\pm\sigma^2\phi)}
\left(\sqrt{3\pi G}\mp2\sigma^2\phi\right)\nn\\
&&\qquad\quad
+\,e^\frac{3\pi G}{2\sigma^2}
\left(k_*^2+\sigma^2+3\pi G\right)
\left[1+\mathrm{erf}\left(\frac{\sqrt{3\pi G}\pm2\sigma^2\phi}{\sqrt{2}\,\sigma}\right)
\right]
\Bigg],\\
\label{eqn:Wpm of phi}
W_\pm(\phi)&=&
\frac{8\left(\Pl^2\gamma\lambda\right)^2}{3\pi G}
\Bigg[
\frac{4\sigma}{\sqrt{2\pi}}\,e^{\mp2\phi(\sqrt{12\pi G}\pm\sigma^2\phi)}
\left(\sqrt{3\pi G}\mp\sigma^2\phi\right)\nn\\
&&\qquad
+\,e^\frac{6\pi G}{\sigma^2}
\left(k_*^2+3\,\sigma^2+12\pi G\right)
\left[1+\mathrm{erf}\left(\frac{\sqrt{6\pi G}\pm\sqrt{2}\,\sigma^2\phi}{\sigma}\right)
\right]
\Bigg],
\ea
which give the evolutions of $\langle\hat{V}\vert_\phi\rangle$ and $(\Delta\hat{V}\vert_\phi)$ by \eqnref{eqn:bouncing sol} and \eqnref{eqn:bouncing sol of V square}. The evolutions of $\langle\hat{V}\vert_\phi\rangle$ and $(\Delta\hat{V}\vert_\phi)^2/\langle\hat{V}\vert_\phi\rangle^2$ are depicted in \figref{fig:evolution}.\footnote{To exaggerate the quantum effect, in the figure, we choose $k_*\gg\sqrt{\pi G}$ and $k_*\gg\sigma$ but \emph{not} $\sigma\gg\sqrt{\pi G}$. As a caveat, the limiting constant of the relative uncertainty spread is given by \eqnref{eqn:limiting relative spread part a} instead of \eqnref{eqn:limiting relative spread part b} and it does not necessarily decreases as $\sigma$ increases as suggested by \eqnref{eqn:limiting relative spread part b}.}

Particularly, we are interested in the limiting constants:
\ba
V_\pm&:=&V_\pm(\pm\infty)
=\frac{4\left(\gamma\Pl^2\lambda\right)e^\frac{3\pi G}{2\sigma^2}}
{\sqrt{3\pi G}k_*}\left(k_*^2+\sigma^2+3\pi G\right),\\
W_\pm&:=&W_\pm(\pm\infty)
=
\frac{16\left(\gamma\Pl^2\lambda\right)^2e^\frac{6\pi G}{\sigma^2}}
{3\pi G}\left(k_*^2+3\sigma^2+12\pi G\right),
\ea
which are used to compute the constant relative uncertainty spread in the WDW regime in terms of $k_*$ and $\sigma$:
\begin{subequations}\label{eqn:limiting relative spread}
\ba\label{eqn:limiting relative spread part a}
\lim_{\phi\rightarrow\pm\infty}
\frac{(\Delta V\vert_\phi)^2}{\langle\hat{V}\vert_\phi\rangle^2}
&=& (W_\pm-V_\pm^2)/V_\pm^2
=
\frac{e^{\frac{3\pi G}{\sigma^2}} k_*^2\left(k_*^2+3\sigma^2+12\pi G\right)}
{\left(k_*^2+\sigma^2+3\pi G\right)^2}-1\\
\label{eqn:limiting relative spread part b}
&\approx& e^{\frac{3\pi G}{\sigma^2}}-1
\approx \frac{3\pi G}{\sigma^2},
\quad\text{if }k_* \gg \sigma \gg \sqrt{\pi G}.
\ea
\end{subequations}

Finally, to know the epoch of the quantum bounce, let us compute
\ba
\partial_\phi\langle\hat{V}\vert_\phi\rangle
&=&V'_+(\phi)\,e^{\sqrt{12\pi G}\,\phi}
+V'_-(\phi)\,e^{-\sqrt{12\pi G}\,\phi}
+\sqrt{12\pi G}\left(V_+(\phi)\,e^{\sqrt{12\pi G}\,\phi}
-V_-(\phi)\,e^{-\sqrt{12\pi G}\,\phi}\right)\nn\\
&=&\frac{16\gamma\Pl^2\lambda}{\sqrt{12\pi G}}
\left(v_+(\phi)\,e^{\sqrt{12\pi G}\,\phi}-v_-(\phi)\,e^{-\sqrt{12\pi G}\,\phi}\right)\nn\\
&&\quad +\sqrt{12\pi G}\left(V_+(\phi)\,e^{\sqrt{12\pi G}\,\phi}
-V_-(\phi)\,e^{-\sqrt{12\pi G}\,\phi}\right).
\ea
It is easy to show that $\partial_\phi\langle\hat{V}\vert_\phi\rangle=0$ if and only if $\phi=0$. Therefore, $\phi=0$ is the epoch of the quantum bounce and the expectation value of the volume at the quantum bounce is given by
\begin{subequations}\label{eqn:Vbounce}
\ba
V_\mathrm{bounce}&=&\langle\hat{V}\vert_{\phi=0}\rangle
=V_+(\phi=0)+V_-(\phi=0)\nn\\
&=&\frac{8\gamma\Pl^2\lambda\,\sigma}{\sqrt{2\pi}k_*}
\left(
1+\frac{e^\frac{3\pi G}{2\sigma^2}}{\sqrt{6G}\,\sigma}
\left(k_*^2+\sigma^2+3\pi G\right)
\left[1+\mathrm{erf}\left(\frac{\sqrt{3\pi G}}{\sqrt{2}\,\sigma}\right)\right]
\right)\\
&\approx&
\frac{8\gamma\Pl^2\lambda\,e^\frac{3\pi G}{2\sigma^2}}{\sqrt{12\pi G}k_*}
\left(k_*^2+\sigma^2\right),
\quad\text{if }k_*, \sigma \gg \sqrt{\pi G}.
\ea
\end{subequations}
By \eqnref{eqn:expectation of pphi}, the matter density at the bouncing epoch is given by
\begin{subequations}\label{eqn:density at bounce}
\ba
\rho_\mathrm{bounce}^{(\infty)}&:=&\frac{\langle\hat{p}_\phi\rangle^2}{2V_\mathrm{bounce}^{\,2}}
=\frac{3\pi^2 e^{-\frac{3\pi G}{\sigma^2}}}{4}\,
\rho_\mathrm{Pl}\\
&\approx& \left(1-\frac{3\pi G}{\sigma^2}\right)\rho_{\sup}^{(\infty)},
\quad\text{if }\sigma\gg\sqrt{\pi G},
\ea
\end{subequations}
where $\rho_{\sup}^{(\infty)}$ is given in \eqnref{eqn:rho sup}.
Therefore, for semiclassical states ($k_*\gg\sigma\gg\sqrt{\pi G}$), $\rho_\mathrm{bounce}^{(\infty)}$ is smaller than but very close to $\rho_{\sup}^{(\infty)}$. The consideration with coherent states shows that on $\hil_\phy$, $\rho_\mathrm{bounce}^{(\infty)}$ can come arbitrarily close to $\rho_{\sup}^{(\infty)}$. A similar result has been shown for the case of $n=0$ in \cite{Ashtekar:2007em}.

\ \newline\noindent
\textbf{Remark:}

One of the virtues to adopt the improved scheme as in \eqnref{eqn:mu bar} is that the resulting dynamics is independent of the elementary cell $\mathcal{V}$ at the level of heuristic effective dynamics, hence giving the correct semiclassical behavior (see \cite{Chiou:2007mg} for more details). In quantum theory of LQC, however, the invariance under the different choice of $\mathcal{V}$ is no longer exact but respected only in the semiclassical regime. This can be seen from \eqnref{eqn:Vpm of phi} and \eqnref{eqn:Wpm of phi}, where the constant $\sqrt{\pi G}$ arises in the company of $k_*$ and/or $\sigma$ inside the parentheses. Since both $k_*$ and $\sigma$ scale linearly with respect to the size of $\mathcal{V}$ while $\sqrt{\pi G}$ is simply constant, the appearance of $\sqrt{\pi G}$ breaks down the independence of $\mathcal{V}$, which now holds only if $k_*, \sigma \gg \sqrt{\pi G}$. As a consequence, the matter density at the bouncing epoch is slightly dependent on $\sigma$ (and thus on $\mathcal{V}$ as well) according to \eqnref{eqn:density at bounce}. It is a common phenomenon that a quantum system reacts to macroscopic scales introduced by boundary conditions; the breakdown of the scaling invariance is reminiscent of the well-known ``conformal anomaly'' as a ``soft'' breaking of conformal symmetry. Also note that the breaking of independence of $\mathcal{V}$ is generic for any order $n$ of holonomy corrections.

\ \newline\noindent
\textbf{Remark:}

Recall that in the $\nu$ representation, the Hamiltonian equation \eqnref{eqn:diffence equation} with higher order holonomy corrections gives rise to a higher order difference equation. It has been argued in \cite{Bojowald:2003dn} that, if the higher order difference equation admits solutions with growing amplitudes, the difference equation is not locally stable. Furthermore, the analysis of \cite{Perez:2005fn} also suggests the existence of spurious solutions in LQG for the Hamiltonian constraint in $j>1/2$ representations. It was shown in \cite{Vandersloot:2005kh} for LQC that the Hamiltonian constraint operator in $j>1/2$ representations indeed allows spurious solutions which eventually break down the semiclassicality in the large scale. The ill-behaved spurious solutions suggested by the earlier studies call into question the validity of the higher $j$ quantization. However, our investigation in this subsection shows that (at least for the case of $n=\infty$) the expectation values of Dirac observables are well behaved and follow the WDW trajectories in the large scale, given that $F(x)$ is a physical state. In other words, if the physical solution comes close to the WDW solution, it will continue following the WDW trajectory in the larger scale. Apparently, the problem of spurious solutions is gone in our analysis; two different explanations might explain the avoidance of spurious states. The first explanation is that the spurious solutions come out only in the quantization of a specific $j$ but are suppressed if all $j$ representations are included to match the desirable form of \eqnref{eqn:Cprime n}, which is motivated to yield better semiclassical behavior. The second possibility is that spurious solutions do exist in the kinematical Hilbert space but have vanishing or infinite physical norms in the physical Hilbert space $\hil_\phy$ and thus are excluded in $\hil_\phy$. It is not clear whether the quantum evolution is still free of spurious solutions if the higher order corrections on the inverse volume operator as in \eqnref{eqn:jV-1} are also taken into account. More studies are awaited for the issues of local stability and spurious solutions.

\section{Summary and discussion}\label{sec:discussion}
By applying the techniques introduced in \cite{Ashtekar:2007em}, we have successfully formulated the quantum theory of LQC for the $k=0$ FRW model with the extension of higher order holonomy corrections.

At any arbitrary order $n$ of holonomy corrections, the physical Hilbert space is rigorously constructed and a complete family of Dirac observables are identified. Two natural notions representing the measurement of the matter density, as defined in \eqnref{eqn:density notion 1} and \eqnref{eqn:rho bounded 2}, are used to compute the expectation value of the matter density. As far as both notions are concerned, for any arbitrary physical states (not only restricted to the states which are semiclassical at late times), the matter density remains bounded from above by an absolute upper bound $\rho_{\sup}^{(n)}$ as shown in \eqnref{eqn:rho bounded 1} and \eqnref{eqn:rho bounded 2}. Therefore, following the same notion of singularity resolution suggested in \cite{Ashtekar:2007em}, the classical singularity is said to be resolved in the quantum theory. We thus extend the key results of \cite{Ashtekar:2007em} to the case of generic $n$. Furthermore, the upper bound $\rho_{\sup}^{(n)}$  given by \eqnref{eqn:rho bounded 1} increases with $n$ but remains finite in the Planckian regime even for $n=\infty$. The value of $\rho_{\sup}^{(n)}$ are exactly the same as that of the critical density $\rho_\mathrm{crit}^{(n)}$ obtained in the heuristic analysis of \cite{Chiou:2009hk}.

Particularly, for the case of $n=\infty$, it is proved that the expectation value of the volume operator $\langle\hat{V}\vert_\phi\rangle$ gives rise to the bouncing scenario in which two WDW solutions are bridged together through a transition phase of the quantum bounce. Given that $F(x)$ is smooth, the trajectories of $\langle\hat{V}\vert_\phi\rangle$, $\langle(\hat{V}\vert_\phi)^2\rangle$, and so on are all smooth functions of $\phi$, despite the kink of the function $b_h(x)=b_h^{(\infty)}(x)$ at $x=0$. Furthermore, the detailed analysis for the coherent states shows that the matter density at the bouncing epoch $\rho_\mathrm{bounce}^{(\infty)}$ is smaller than but can come arbitrarily close to $\rho_{\sup}^{(\infty)}$.

On the other hand, we also explore the idea that the higher order holonomy corrections can be interpreted as generic $j$ representations for holonomies in the Hamiltonian constraint operator. We demonstrate that it might be possible to have a linear superposition of the Hamiltonian operators in generic $j$ representations match the well-motivated operator $\hat{C}^{\prime(n=\infty)}_\grav$. However, the coefficients $c_j^{(n)}$ in \eqnref{eqn:linear superposition} diverge as $n\rightarrow\infty$ and need to be regulated. The peculiar feature of the divergence of $c_j^{(n)}$ is reminiscent of the infrared divergence encountered in many spin-foam models \cite{Ponzano-Regge,Freidel:2002dw,Baez:1999sr} and the regularization may correspond to a nonzero cosmological constant \cite{Noui:2002ag}. We hope that our study inspires further research on the issues of $j$ ambiguity by exploring the link between LQC and the spin-foam formalism. For example, the problem of finding a crossing-symmetric linear combination of the Hamiltonian operators in LQG as proposed in \cite{Gaul:2000ba} could be related to that of finding the linear sum of $\hat{C}^{\prime(n)}_\grav$ to match the operator $\hat{C}^{\prime(n=\infty)}_\grav$ in LQC.

Finally, it should be remarked that the problem of spurious states associated with the higher $j$ representation as suggested in \cite{Vandersloot:2005kh,Perez:2005fn} does not seem to happen (at least for the case of $n=\infty$) in our model. However, we do not know whether it is because the spurious solutions are suppressed in the kinematical Hilbert space when all $j$ representations are properly summed or because they have zero or infinite physical norms and thus are excluded in the physical Hilbert space.

\begin{acknowledgements}
The authors would like to thank Xiangdong Zhang for useful discussions, which helped to initiate this work.
D.W.C. is supported by the NSFC Grant No. 10675019 and the financial support No. 20080440017 from China Postdoctoral Science Foundation; L.F.L. is supported by the NSFC Grant No. 10875012.
\end{acknowledgements}


\end{document}